\documentclass[10pt]{article} 
\usepackage[preprint]{rlc}
\usepackage{authblk} 
\usepackage{booktabs}
\usepackage{tabularx} 
\usepackage{array}
\usepackage{amssymb}            
\usepackage{mathtools}          
\usepackage{mathrsfs}           
\mathtoolsset{showonlyrefs}     
\usepackage{graphicx}           
\usepackage{subcaption}         
\usepackage[space]{grffile}     
\usepackage{listings}           
\usepackage{xurl}               
\usepackage{xspace}             
\usepackage{amsmath}
\usepackage{bm}
\usepackage{multirow}
\usepackage{enumitem}
\usepackage{float}
\usepackage{needspace}
\usepackage{placeins}
\usepackage{wrapfig}  
\usepackage[linesnumbered,ruled,vlined]{algorithm2e}
\usepackage{xcolor}
\setcitestyle{numbers,square,sort&compress,citesep={,}}
\hypersetup{citecolor={black}}
\newcommand{\reviewchange}[1]{#1}
\newif\iftokenrevisionmarkup
\tokenrevisionmarkupfalse
\newcommand{\tokenrevision}{\iftokenrevisionmarkup\color{red}\fi}
\setlist[itemize]{
    leftmargin=17pt
}

\makeatletter
\newcommand{\ie}{\emph{i.e.}\@ifnextchar.{\!\@gobble}{}}
\newcommand{\eg}{\emph{e.g.~}\@ifnextchar.{\!\@gobble}{}}
\newcommand{\etc}{etc\@ifnextchar.{}{.\@}}

\newcommand{\Duo}{\textsc{Duo}}
\newcommand{\Trio}{\textsc{Trio}}
\newcommand{\Quad}{\textsc{Quad}}
\newcommand{\maximize}{\mathop{\mathrm{maximize}}}

\makeatother

\title{MARFT: Multi-Agent Reinforcement Fine-Tuning}

\author{
    \name{Junwei Liao}\addr{$^{1,2}$}, \name{Muning Wen}\addr{$^{1}$}, \name{Jun Wang}\addr{$^{3}$}, \name{Weinan Zhang}\addr{$^{1,2}$}\\
    \addr{$^1$Shanghai Jiao Tong University}, \addr{$^2$Shanghai Innovation Institute}, \addr{$^3$OPPO Research Institute}\\
    \email{\{jwliao.ai, muningwen, wnzhang\}@sjtu.edu.cn}, \email{\{junwang.lu\}@gmail.com}
    \\
}

\begin{document}

\maketitle

\begin{abstract}
Large Language Model (LLM)-based Multi-Agent Systems (LaMAS) have demonstrated remarkable capabilities in addressing complex, agentic tasks requiring multifaceted reasoning and collaboration, from generating high-quality presentation slides to conducting sophisticated scientific research. Meanwhile, Reinforcement Learning (RL) has been widely recognized for its effectiveness in enhancing agent intelligence, but limited research has investigated the fine-tuning of LaMAS using foundational RL techniques. Moreover, the direct application of conventional Multi-Agent Reinforcement Learning (MARL) methodologies to LaMAS introduces significant challenges, stemming from the unique characteristics and mechanisms inherent to LaMAS. To address these challenges, this article presents a comprehensive study of LLM-based MARL and proposes a novel paradigm termed \textbf{Multi-Agent Reinforcement Fine-Tuning (MARFT)}. We introduce a brand-new MG called Flex-MG, which aligns with the LaMAS optimization in real-world applications and a universal algorithmic framework tailored specifically for LaMAS, outlining the conceptual foundations, key distinctions, and practical implementation strategies. We begin by reviewing the evolution from traditional RL to Reinforcement Fine-Tuning (RFT), setting the stage for a parallel analysis in the multi-agent domain. In the context of LaMAS, we elucidate critical differences between classical MARL and MARFT, such as asynchronous agent interactions, profile-aware agent design, and heterogeneous architectural configurations etc. These differences motivate a transition toward a novel, LaMAS-oriented formulation of RFT. Central to this work is the presentation of a robust and scalable MARFT framework. We detail the core algorithm, emphasizing its modularity and adaptability, and provide a complete, open-source implementation to facilitate adoption and further research. Experiments on fixed two- to four-agent mathematical and coding workflows compare MARFT with and without policy parameter sharing against IPPO: both variants report higher mean evaluation scores in all 24 benchmark--team-size settings, while role-prompting ablations examine their training behavior. The latter sections of the paper explore real-world application perspectives and opening challenges in MARFT, including dynamic environment modeling, sample inefficiency, and the current lack of cohesive frameworks. By bridging theoretical underpinnings with practical methodologies, this work aims to serve as a roadmap for researchers seeking to advance MARFT toward resilient, adaptive, and human-aligned solutions in agentic systems. Our implementation of the proposed framework is publicly available at: \href{https://github.com/jwliao-ai/MARFT}{https://github.com/jwliao-ai/MARFT}.
\end{abstract}

 \section{Introduction} 
\label{sec:introduction}

Large Language Models (LLMs) are increasingly being deployed as a new generation of autonomous agents capable of performing agentic tasks—those that require decision-making, reasoning, and interaction with complex and dynamic environments \citep{jin2024llmsllmbasedagentssoftware,hong2024metagpt,qian-etal-2024-chatdev}. These LLM-based agents are rapidly reshaping human–machine interaction and expanding the frontiers of autonomous systems. In addition to their strong natural language understanding and generation capabilities \citep{Chowdhary2020}, LLMs can perform retrieval-augmented generation (RAG) \citep{lewis2021retrievalaugmentedgenerationknowledgeintensivenlp}, and, when integrated with external tools or APIs, can accomplish more sophisticated tasks on computers and mobile platforms \citep{erdogan2024tinyagentfunctioncallingedge,zhang2025agenticinformationretrieval}. Furthermore, LLMs have been successfully embedded into embodied and simulated environments, functioning as agents such as robots or game players \citep{tan2024true,mandi2023roco,pmlr-v202-carta23a}. Their ability to comprehend instructions, learn from feedback, and generate context-aware outputs enables effective deployment across various domains, including healthcare, education, and software development \citep{dai2023adautogptautonomousgptalzheimers,chen2024octopusondevicelanguagemodel}.

Reinforcement Learning (RL) is a machine learning paradigm in which an agent learns optimal behaviors through trial-and-error interactions with an environment to maximize cumulative reward. Unlike supervised or unsupervised learning, RL is driven by feedback in the form of rewards or penalties \citep{sutton1998introduction}. The recent release of OpenAI's o1 model, trained using large-scale RL techniques, has reignited interest in the field due to its impressive reasoning abilities \citep{openai2024learningtoreason}. Following this, OpenAI expanded its broader reinforcement fine-tuning (RFT) research program focused on refining LLMs through RL approaches \citep{openai2024rftresearchprogram}. Distinct from conventional RL, RFT involves fine-tuning pretrained models using a relatively small set (dozens to thousands) of high-quality interaction trajectories. This process must maintain the language capabilities of the LLM while enhancing its reasoning and decision-making skills. Recent studies have shown that RFT can significantly improve the performance of LLM-based agents \citep{deepseekai2025deepseekr1incentivizingreasoningcapability,qwen2025qwq32b,shao2024deepseekmathpushinglimitsmathematical,zeng2024scalingsearchlearningroadmap,wang2024openropensourceframework}. \reviewchange{Notably, GRPO was proposed as a variant of Proximal Policy Optimization (PPO), enhancing reasoning abilities and revealing intriguing phenomena such as the ``Aha moment'' during large-scale RL training \citep{shao2024deepseekmathpushinglimitsmathematical}. Building on this, DAPO and VAPO extend GRPO with additional techniques \citep{yu2025dapoopensourcellmreinforcement,yue2025vapoefficientreliablereinforcement}.} Collectively, these efforts highlight the centrality of the trial-and-error nature of RL in advancing LLM intelligence. Despite these advances, directly applying RFT to multi-agent LLM settings remains underexplored. Naively transferring single-agent RFT techniques to multi-agent contexts often leads to challenges such as unstable training, inactive agents, and inefficient communication. These issues underscore the need to revisit and adapt established concepts from MARL to bridge the gap between RFT and LLM-based Multi-Agent Systems (LaMAS).

Multi-Agent Systems (MAS) have consistently demonstrated superior capabilities in addressing complex agentic tasks relative to their single-agent counterparts. This is evident in the latest General AI Assistants (GAIA) leaderboard \citep{mialon2024gaia}, where the top-performing systems are all multi-agent frameworks. Prominent frameworks, including OpenAI Agents SDK (formerly Swarm), Microsoft AutoGen \citep{wu2023autogenenablingnextgenllm}, Magnetic-One, CAMEL-AI OWL \citep{owl2025}, and Google's AI Co-Scientist \citep{gottweis2025aicoscientist}, further reinforce the growing importance of multi-agent architectures. Notably, in OpenAI's vision for AGI, multi-agent systems constitute the highest organizational layer. \reviewchange{However, multi-agent systems are prone to a variety of failure modes--fourteen in total--related to system design, coordination, and control, where MARL is poised to address these challenges \citep{cemri2025multiagentllmsystemsfail}. Besides, a recent framework systematically quantifies the contributions of different modules in multi-agent systems, potentially illuminating the direction for future multi-agent optimization in LLMs \citep{yang2025whosmvpgametheoreticevaluation}.}

Multi-Agent Reinforcement Learning (MARL) extends standard RL to settings where multiple autonomous agents interact and learn simultaneously to achieve individual or shared objectives \citep{weiss1999multiagent,busoniu2008comprehensive,stone2000multiagent,tuyls2012multiagent}. Applications of MARL span numerous domains, including swarm robotics \citep{JMLR:v20:18-476,Matignon_Jeanpierre_Mouaddib_2021}, autonomous driving \citep{shalevshwartz2016safemultiagentreinforcementlearning}, strategic games \citep{song2024empirical}, traffic management \citep{6958095}, and distributed resource allocation \citep{9563249}. MARL methods can be broadly categorized into independent learning, joint learning, and coordination-based strategies \citep{10.5555/3091529.3091572,Matignon_Laurent_LeFort-Piat_2012,1470239}. A widely adopted paradigm is centralized training with decentralized execution (CTDE), with prominent algorithms including MADDPG \citep{lowe2017multi}, MAPPO \citep{yu2022the}, and HAPPO \citep{kuba2022trust}. \reviewchange{More recently, Multi-Agent Transformer (MAT) reframes MARL as a series of sequential decision-making problems, potentially opening new directions for research \citep{MAT}.} Despite these advances, integrating LLMs into MARL settings introduces significant challenges. Unlike conventional multi-agent systems in traditional MARL tasks, LaMAS are highly flexible and complex, particularly in their communication dynamics. Addressing the challenges requires rethinking fundamental assumptions in MARL, such as agent homogeneity and communication protocols, to accommodate the unique properties of LaMAS. Though there are easy-to-use and well-established frameworks such as MALib \citep{JMLR:v24:22-0169} unifying many existing MARL methods for population-based RL and OpenRLHF \citep{hu2024openrlhf} providing implementations for most single-agent RL fine-tuning techniques, the absence of a unified framework that integrates LLMs as agents in dynamic environments with agentic tasks still constrains the swarm intelligence of LaMAS.
Consequently, there remains a need for a well-established LLM-based multi-agent framework both theoretically and practically to unleash LaMAS's swarm intelligence and potential better. 

\begin{figure}
    \centering
    \includegraphics[width=0.99\linewidth]{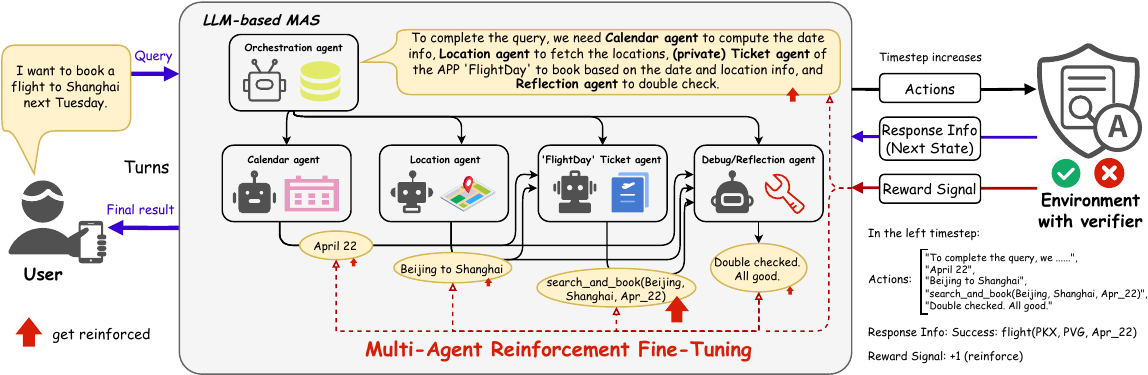}
    \caption{Illustration of MARFT in real-world agentic problem-solving scenarios.}
    \label{fig:illustration-marft-real-world-scenarios}
\end{figure}

Given the accelerating progress in both RFT for LLMs and LaMAS, we are motivated to introduce and formalize the concept of \textbf{Multi-Agent Reinforcement Fine-Tuning (MARFT)}. This article aims to provide a comprehensive foundation for MARFT, serving both academic and industrial audiences. The remainder of this article is organized as follows: Section~\ref{sec:review} revisits foundational concepts in RFT, LaMAS, and MARL. Section~\ref{sec:overview} gives the problem statement and formulation and outlines the key distinctions between MARFT and existing approaches. Section~\ref{sec:marft} introduces the MARFT methodology and its basic implementation strategies. \reviewchange{Section~\ref{sec:experiments} evaluates fixed two- to four-agent workflows on DeepScaleR and DeepCoder Preview, comparing MARFT with and without policy parameter sharing against IPPO and analyzing agent count and role prompting. Appendix~\ref{app:experiment-hyperparameters} reports the optimization settings, evaluation protocol, and system prompts.} Finally, Sections~\ref{sec:perspectives} and~\ref{sec:open-problems} provide future perspectives and highlight unresolved challenges, followed by an overall conclusion of the paper in Section~\ref{sec:conclusion}.

\section{Literature Review}
\label{sec:review}

This section presents a comprehensive review of topics closely related to MARFT, namely RFT, LaMAS, and MARL. We begin with a discussion of the algorithmic foundations and methodological formulations of RL, followed by an in-depth review of RFT. Subsequently, we provide an overview of LaMAS and MARL from a broader perspective, thereby laying the conceptual groundwork for the introduction and development of MARFT.

\subsection{Reinforcement Fine-Tuning (RFT)}
\label{sec:review:rft}

RFT is a fine-tuning paradigm that harnesses the capabilities of Reinforcement Learning to refine agent behavior. While it shares foundational similarities with traditional RL, RFT places greater emphasis on achieving expert-level performance in specific tasks using limited, high-quality data and a constrained number of optimization iterations. To provide context for the discussion of RFT, we first briefly revisit the core principles, mathematical formulations, and algorithmic methodologies of standard RL.

\subsubsection{Language-augmented Markov Decision Process}
\label{sec:review:rft:language-augmented-mdp}
In most sequential decision-making scenarios, problems are typically modeled as a \textit{Markov Decision Process (MDP)} $\langle\mathcal{S},\mathcal{A},\mathcal{T},\mathcal{R}, \gamma\rangle$. In linguistic contexts, this framework can be further extended to a more specific form of language-augmented \textit{Partial Observable Markov Decision Process (POMDP)} represented as $\langle\mathcal{V}, \mathcal{S},\mathcal{O},\mathcal{A},\mathcal{T},\mathcal{R},\gamma\rangle$ \citep{vanOtterlo2012,pmlr-v202-carta23a}. Here, $\mathcal{V}$ represents the vocabulary of a language model, which facilitates natural language understanding within an LLM. The state space $\mathcal{S}$ can be situated in a semantic space, particularly in linguistic contexts. The observation function $\mathcal{O} \colon \mathcal{S} \rightarrow \mathcal{V}^L$ and action space $\mathcal{A}\subseteq\mathcal{V}^L$ respectively denote the observations and actions, both of which are explicitly represented by sequences of $L$ tokens from the vocabulary $w \in \mathcal{V}$. Furthermore, the action $a\in\mathcal{A}$ can be grounded to a specific tool or API call in real-world environments \citep{10.5555/3666122.3669119,berkeley-function-calling-leaderboard, lu2024toolsandboxstatefulconversationalinteractive}. The transition function $\mathcal{T}\colon\mathcal{S}\times\mathcal{A}\rightarrow\mathcal{S}^\prime$, often represented as a probability distribution $\mathcal{P}$ in the form of $P(s^\prime|s, a)$ describes the state transition following a specific action taken in a given state. However, in linguistic settings, the transition for observations can be directly achieved through sentence concatenation. The reward function $\mathcal{R}\colon\mathcal{S}\times\mathcal{A}\times\mathcal{S}\rightarrow\mathbb{R}$, typically denoted as $R(s, a, s^\prime)$, quantifies the reward for taking a particular action that induces a state transition. The discount factor $\gamma$ reflects the diminishing value of rewards received over time. This constitutes the fundamental language-augmented POMDP, where a single LLM takes actions within a language-enhanced environment. For any other language-based tasks, specific POMDPs can be derived from this basic language-augmented POMDP.

\subsubsection{Reinforcement Learning (RL)}
\label{sec:review:rft:rl}

\paragraph{The Objective of RL} In classic reinforcement learning settings, an agent is capable of perceiving its environment, making decisions, and taking actions, while learning from past experiences to enhance its performance. The agent interacts with the environment until certain terminal conditions are met, generating a trajectory $\tau = (s_0, a_0, r_0, s_1, a_1, r_1, \cdots, s_T, a_T, r_T)$ of length $T$, with a cumulative reward expressed as $g = \sum_{t=0}^T\gamma^tr_t$. The primary objective is for the agent to achieve a high cumulative reward in a dynamic and uncertain environment. This is precisely what RL aims to accomplish—obtaining a policy that enables the agent to dynamically act in dynamic environments to maximize cumulative reward. Mathematically, the objective is formulated as $\max\limits_\pi \mathbb{E}_{\tau\sim p_\pi(\tau)}\left[\sum_{t=0}^T\gamma^tR_t\right]$ \citep{sutton1998introduction}. Here, the cumulative reward from timestep $t$ is denoted as $G_t = \sum_{t^\prime=t}^\infty\gamma^{t^\prime-t}R_{t^\prime}$.

\paragraph{Value-based Methods} Given that RL optimizes the policy by leveraging rewards to incentivize the policy to learn actions that yield the highest rewards, it is natural to consider a value-based approach. This value-based approach provides an elegant and effective solution to the critical credit assignment problem \citep{4066245,sutton1998introduction}. This approach focuses on learning indirect rewards, i.e, values, which estimate the goodness of an action before receiving the reward and ideally can better optimize the policy in long-horizon and reward-sparse tasks. To this end, two vital proxies to the RL objective are introduced: the state-action value function (Q-function) and the state value function (V-function) \citep{sutton1998introduction}. Their definitions are presented respectively as
\begin{gather}
    \label{eq:value-functions}
    Q^\pi(s_t, a_t) \coloneqq \mathbb{E}_{\tau \sim p_\pi(\tau|s_t, a_t)}\left[\sum_{t^\prime=t}^\infty \gamma^{t^\prime-t}r(s_{t^\prime}, a_{t^\prime})\right] = \mathbb{E}_{(s,a)_{t+1:\infty} \sim p_\pi(s,a)}\left[G_t|s_t, a_t \right],\\
    V^\pi(s_t) \coloneqq \mathbb{E}_{\tau \sim p_\pi(\tau|s_t)}\left[\sum_{t^\prime=t}^\infty \gamma^{t^\prime-t}r(s_{t^\prime}, a_{t^\prime})\right] = \mathbb{E}_{a_t \sim \pi(a_t|s_t), (s,a)_{t+1:\infty} \sim p_\pi(s,a)}\left[G_t|s_t\right].
\end{gather}
Here $p_\pi$ is the trajectory distribution sampled using the policy $\pi$. Obviously, from their naming and mathematical expressions, it is evident that the Q-function indicates how valuable an action $a_t$ is at state $s_t$, while the V-function indicates the value of being in state $s_t$. Another important proxy function is the advantage function, defined as $A_\pi(s_t, a_t) \coloneqq Q_\pi(s_t, a_t) - V_\pi(s_t)$ \citep{NIPS2001_584b98aa}. This function quantifies how much better or worse an action $a_t$ is compared to the average action at state $s_t$.

By expanding the definitions and expressing the summation terms using $Q$ or $V$, we can derive two recursive equations as 
\begin{gather}
    \label{eq:q-v-dual-express}
    Q^\pi(s_t, a_t) = r(s_t, a_t) + \gamma \mathbb{E}_{s_{t+1}\sim p(s_{t+1}|s_t, a_t)}\left[V^\pi(s_{t+1})\right],\\
    V^\pi(s_t) = \mathbb{E}_{a_t\sim \pi(a_t|s_t)}\left[Q^\pi(s_t, a_t)\right].
\end{gather}
By substituting these two equations into each other, we can further derive the Bellman equations \citep{kirk1970optimal} as
\begin{gather}
    \label{eq:bellman-equations}
    Q^\pi(s_t, a_t) = r(s_t, a_t) + \gamma \mathbb{E}_{s_{t+1}\sim p(s_{t+1}|s_t, a_t), a_{t+1}\sim \pi(a_{t+1}|s_{t+1})}\left[Q^\pi(s_{t+1},a_{t+1})\right],\\
    V^\pi(s_t) = \mathbb{E}_{a_t\sim \pi(a_t|s_t)}\left[r(s_t, a_t) + \gamma \mathbb{E}_{s_{t+1}\sim p(s_{t+1}|s_t, a_t)}\left[V^\pi(s_{t+1})\right]\right].
\end{gather}
There are two primary methods for estimating value functions. For clarity, we focus on the V-function here, as the estimation of the Q-function can be similarly derived. The most straightforward approach is to use Monte Carlo (MC) estimation \citep{sutton1998introduction}. By performing multiple Monte Carlo samplings, we can approximate the value function as: $V(s_t)\approx\frac{1}{K}\sum_{k=1}^K\sum_{i=t}^\infty\gamma^{i-t}r\left(s_i^k, a_i^k\right)$. According to the law of large numbers \citep{ref1}, this method provides an unbiased estimation of the value function. However, it can be highly inefficient, costly, and time-consuming, especially when the agent is an LLM.

A more preferable approach is to use Temporal Difference (TD) learning \citep{10.5555/911176}. The value Bellman equation, which resembles dynamic programming, bootstraps the value function over itself at the next timestep. Therefore, we can perform Bellman backups to estimate the value function. In the context of deep learning, we typically construct a value function (also known as a critic network) \citep{mnih2013playingatarideepreinforcement} and train it by minimizing the empirical Bellman error as
\begin{align}
    \label{eq:bellman-error-loss}
    L(\phi) = \frac{1}{|\mathcal{D}|} \sum_{\left(o_t,a_t,o_{t+1},r_t\right)\sim \mathcal{D}} \left[ r_t(o_t, a_t) + \gamma V_{\phi}(o_{t+1}) - V_{\phi}(o_t) \right]^2.
\end{align}
Once the value function is obtained, the next crucial step is policy extraction, which involves deriving the policy from the value function \citep{Watkins1992}. The specific policy extracted can vary widely depending on the chosen method. For instance, a greedy policy can be extracted using $\pi\left(a_t|s_t\right) = \delta\left(a_t = \mathrm{argmax}Q\left(s_t, a_t\right)\right)$, where $\delta$ denotes the Dirac delta function. Alternatively, a soft policy can be derived as $\pi_{\text{MaxEnt}}\left(a_{t}|s_{t}\right) = \exp \left( \frac{1}{\alpha} \left( Q_{\text{soft}}\left(s_{t}, a_{t}\right) - V_{\text{soft}}\left(s_{t}\right) \right) \right)$, where $\alpha$ is a temperature parameter \citep{haarnoja2017reinforcement}. Other extraction methods can yield different policies, including mixture policies. Algorithm \ref{alg:general-value-based-method} depicts the fundamental procedure of value-based deep RL methods. However, policy extraction in the context of LLMs can be highly counterintuitive. We will elaborate on the reasons for this in Section \ref{sec:overview:differences-and-challenges}.

\begin{algorithm}[htb]
    \caption{General value-based deep RL method}
    \label{alg:general-value-based-method}
    \SetAlgoLined
    \KwIn{Parameters $\phi_0$ for value function $Q$, some policy derived from value function \eg $\epsilon$-greedy policy $\pi_0(a|s) = \epsilon U(a) + (1-\epsilon)\delta\left(a =\arg\max_aQ_{\phi_0}\left(s, a\right)\right)$, some initial state distribution $d\left( s_0\right)$, transition probability $p\left(s^\prime|s, a\right)$, discount factor $\gamma$, learning rate $\alpha$, etc.}
    \KwOut{Optimal Q function $Q^\star$ with parameters $\phi^\star$, extracted optimal policy $\pi^\star$}
    initialize parameters $\phi_0$ for value function $Q_\phi$, target value function w/o gradient $Q_{\text{targ}}\gets Q_{\phi_0}$, some policy $\pi_0$, replay buffer $\mathcal{D} \gets \emptyset$\\
    \For{$\text{iterations} \ i=0, \dots$}{
        initialize $s \gets s_0 \sim d(s_0)$\\
        \While{not terminal state}{
            select action using policy $ a \sim \pi_i(a|s)$\\
            observe next state $s^\prime \sim p\left(s^\prime|s, a\right)$\\
            get reward $r = R(s, a)$\\
            $\mathcal{D}\gets\mathcal{D}\cup\{\left(s, a, s^\prime,r \right) \}$\\
        }
        sample a batch $\mathcal{B}=\{(s,a,s^\prime, r)\} \sim \mathcal{D}$\\
        compute Bellman error $\bm\epsilon = \sum_\mathcal{B}\big(Q_{\phi_i}(s,a) - (r+\gamma\max_{a^\prime} Q_{\text{targ}}(s^\prime,a^\prime) \big)$\\
        update $\phi$ by minimizing $\bm\epsilon$: $\phi_{i+1} \gets \phi_i - \alpha\nabla\bm\epsilon$\\
        extract policy $\pi_{i+1}$ based on $Q_{\phi_{i+1}}$\\
        periodically update target value function: $Q_{\text{targ}} \gets Q_{\phi_i}$\\
    }
\end{algorithm}

\paragraph{Policy-based Methods} Another approach to optimizing policies in RL involves directly estimating the gradient of the policy and performing gradient ascent. This method is largely based on the policy gradient (PG) theorem \citep{NIPS1999_464d828b}. Recall that the objective is to maximize the expected reward: $J(\theta) = \mathbb{E}_{_{\tau \sim p_{\pi_{\theta}}(\tau)}} \left[ \sum_{t=0}^{\infty} \gamma^{t} r(s_{t}, a_{t}) \right]$.\footnote{In other papers you may see people denote it $J(\pi_\theta)$ In fact, these two expressions are the same. You can view it as some reward $J$ being a function of parameters $\theta$ or some reward $J$ being a function of the policy $\pi$, which is parameterized by $\theta$.} According to the policy gradient theorem, the gradient of the expected reward $J(\theta)$ can be expressed as 
\begin{align}
    \label{eq:policy-gradient}
    \nabla_{\theta} J(\theta) = \mathbb{E}_{\tau \sim p_{\pi_{\theta}}(\tau)} \left[ \sum_{t=0}^{\infty} \Psi_{t} \nabla_{\theta} \log \pi_{\theta}(a_{t}|s_{t}) \right],
\end{align}
where, in the simplest case, $\Psi_{t} = \sum_{t^\prime=t}^\infty\gamma^{t^\prime - t}r(s_{t^\prime}, a_{t^\prime})$. However, the cumulative discounted reward is often difficult to obtain directly, so $\Psi_t$ can be approximated using other proxies, such as the Q-value $Q^\pi(s,a)$ or the advantage function $A^\pi(s_t, a_t)$ \citep{schulman2018highdimensionalcontinuouscontrolusing}. 

Given the gradient of $J$ with respect to the policy parameters $\theta$, we can do gradient ascent to update the parameters as \eqref{eq:policy-gradient-ascent}. The most classic method is REINFORCE \citep{10.1007/BF00992696}, which utilizes MC sampling to approximate the gradient as 
\begin{equation}
\begin{aligned}
    \nabla_{\theta} J(\theta) &\approx \frac{1}{K} \sum_{k=1}^{K} \sum_{t=0}^{T} \nabla_{\theta} \log \pi_{\theta}(a_{t}^{k}|s_{t}^{k})G_{t}^{k},\\
    \theta & \leftarrow \theta + \alpha \nabla_{\theta} J(\theta).
\end{aligned}\label{eq:policy-gradient-ascent}
\end{equation}
Algorithm \ref{alg:general-policy-based-method} illustrates the general process of policy-based deep RL methods. And most of the time, people subtract the cumulative reward with a baseline $b(s_t)$ defined as an average over the sampled trajectories to reduce the variance \citep{NIPS2001_584b98aa}.

As we can see, policy gradient does not require a value function anymore, instead, it uses the true rewards to optimize the policy directly, which is more stable compared with value-based methods, especially when the policy is an LLM. Moreover, many other variants have been proposed, \eg REINFORCE++ \citep{hu2025reinforcesimpleefficientapproach}, RLOO \citep{kool2019buy}, etc., and they seem to work relatively well in LLMs.

\begin{algorithm}[htb]
    \caption{General policy-based deep RL method}
    \label{alg:general-policy-based-method}
    \SetAlgoLined
    \KwIn{intial parameters $\theta_0$ for policy $\pi_\theta$, initial state distribution $d\left( s_0\right)$, transition probability $p\left(s^\prime|s, a\right)$, discount factor $\gamma$, learning rate $\alpha$, etc.}
    \KwOut{Optimal policy $\pi^\star$}
    initialize parameters $\theta_0$ for policy $\pi_\theta$, replay buffer $\mathcal{D} \gets \emptyset$\\
    \For{$\text{iterations}\ i=0,\dots$}{
        sample intial states $\bm{s_0}\sim d\left(s_0\right)$ and generate trajectories $\bm\tau = \{\tau^k\}=\{(s_0,a_0,\dots,s_H,a_H)^k\}$ using current policy $\pi_{\theta}$\\
        compute cumulative returns $G_{t}^k = \sum_{t^\prime=t}^H \gamma^{t^\prime-t} r_{t^\prime}^k(s_{t^\prime}^k,a_{t^\prime}^k)$ (or some other proxy) for all timesteps $t$\\
        estimate policy gradient $\nabla_{\theta} J(\theta) \approx \frac{1}{K} \sum_{k=1}^{K} \sum_{t=0}^{T} \nabla_{\theta} \log \pi_{\theta}(a_{t}^{k}|s_{t}^{k})G_{t}^{k}$\\
        update policy by gradient ascent $\theta \gets \theta + \alpha \sum_{t=0}^T \nabla_\theta J(\theta)$
    }
\end{algorithm}

\paragraph{Actor-Critic} Actor-Critic (AC) \citep{sutton1998introduction,NIPS1999_6449f44a,6315022,pmlr-v48-mniha16} architecture is a hybrid way, incorporating both value-based methods and policy-based methods, to optimize the policy. While value-based methods extract policy poorly and policy-based methods lack direct estimations to values with a result of low efficiency, Actor-Critic inherits fine-grained credit assignment from value-based methods and optimization stability from policy-based methods. Actor-Critic algorithms train the value function and policy iteratively. For each iteration, it updates the value function based on TD error and optimizes the policy via the policy gradient method

One of the most suitable AC algorithms that adapts well to LLMs is proximal policy optimization (PPO) \citep{schulman2017proximalpolicyoptimizationalgorithms}, which is developed upon Trust Region Policy Optimization (TRPO) \citep{pmlr-v37-schulman15}. TRPO gives a surrogate objective function to maximize, together with a constraint on policy update steps as
\begin{align}
    \label{eq:trpo-objective-with-constraint}
    \maximize_{\theta} \hat{\mathbb{E}}_t \left[ \frac{\pi_{\theta}(a_t|s_t)}{\pi_{\theta_{\text{old}}}(a_t|s_t)} \hat{A}_t(s_t, a_t) \right], \ \mathrm{s.t.} \ \hat{\mathbb{E}}_t \left[ \text{KL} \left[ \pi_{\theta_{\text{old}}}(\cdot|s_t) \| \pi_{\theta}(\cdot|s_t) \right] \right] \leq \delta,
\end{align}
where $\hat{A}_t$ is an estimator of the advantage function at timestep $t$, which is usually computed using GAE \citep{schulman2018highdimensionalcontinuouscontrolusing}, and $\hat{\mathbb{E}}_t$ is the empirical average over a finite batch of samples. By the Lagrange Multiplier, it is equivalent to the unconstrained problem as 
\begin{align}
    \label{eq:trpo-objective-without-constraint}
    \maximize_{\theta} \hat{\mathbb{E}}_t \left[ \frac{\pi_{\theta}(a_t|s_t)}{\pi_{\theta_{\text{old}}}(a_t|s_t)} \hat{A}_t(s_t,a_t) -\beta \text{KL} \left[ \pi_{\theta_{\text{old}}}(\cdot|s_t) \| \pi_{\theta}(\cdot|s_t) \right] \right].
\end{align}
TRPO motivates conservative policy updates through a monotonic-improvement analysis under its stated assumptions, while its practical constrained optimization involves second-order computations, including the Fisher Information Matrix (FIM), which can be costly. PPO provides a first-order clipped surrogate inspired by this trust-region approach; finite-sample clipped PPO does not generally inherit an unconditional monotonic-improvement guarantee. PPO proposes a clipped surrogate object as 
\begin{align}
    \label{eq:ppo-objective}
    \maximize_{\theta} \hat{\mathbb{E}}_{t}\left[ \min \left( \frac{\pi_{\theta}(a|s)}{\pi_{\theta_\mathrm{old}}(a|s)} A^{\pi_{\theta_\mathrm{old}}}(s, a),\text{clip} \left( \frac{\pi_{\theta}(a|s)}{\pi_{\theta_\mathrm{old}}(a|s)}, 1 \pm \epsilon \right) A^{\pi_{\theta_\mathrm{old}}}(s, a) \right)\right].
\end{align}
And countless experiments have shown that PPO is one of the most effective RL methods for nearly all decision-making tasks, though it still suffers from inefficiency and long training time due to the on-policy method and the nature of trial-and-error \citep{10762021,ouyang2022traininglanguagemodelsfollow,heess2017emergencelocomotionbehavioursrich}. 

In LLMs, we can easily analogize from typical RL tasks, \eg, mujoco \citep{todorov2012mujoco}, to language tasks. For example, an LLM is trying to complete the query a user gives, and the LLM generates actions once it receives observations, which can be the user's further responses or function call results, also maybe together with a reward signal. Based on the interaction, or in RL, i.e., trajectory, PPO or other methods can be utilized to reinforce the LLM to learn how to better tackle human queries in interactions \citep{zhou2025sweetrltrainingmultiturnllm,goldie2025syntheticdatageneration}.

\subsubsection{From RL to RFT}
\label{sec:review:rft:from-rl-to-rft}
\begin{table}[htb]
    \caption{Differences between RL and RFT.}
    \label{tab:differences-between-rl-and-rft}
    \begin{center}
        \begin{tabular}{lll}
            \toprule
            \multicolumn{1}{l}{\bf } &\multicolumn{1}{l}{\bf RL} &\multicolumn{1}{l}{\bf RFT} \\
            \midrule
            Actor &Policy from scratch & Large-scale pretrained foundation model\\
            Constraint &No constraint &Maintain original capabilities \\
            Action Space Size &Typically small, numerable &Huge, $O(\mathcal{V}^{|L|})$ or $O(\mathcal{V}\times{|L|})$ \\
            Transition &Stochastic &Deterministic and Stochastic \\
            Granularity  &Action &Action or token \\
            Process  &Reset state and sample traj. &Reset query, sample traj. and verify answer\\
            Sample Magnitude &Large &Relatively small and high-quality \\
            \bottomrule
        \end{tabular}
    \end{center}
\end{table}

\paragraph{Differences} 
Though RL and RFT share foundational methodological principles, there still exist nuanced differences as listed in Table \ref{tab:differences-between-rl-and-rft}. RL, closely rooted in optimal control frameworks, typically trains policy agents, such as robots, from scratch, requiring no prior domain knowledge. In contrast, RFT operates on large-scale pretrained models and adapts these models to specialized tasks. Additionally, RL trains a policy specifically for a particular task within a specific environment, while RFT enhances a model's task-specific capabilities without degrading its original abilities, such as language understanding or generative skills in LLMs. The environments also differ significantly. Conventional RL environments are entirely stochastic, whereas typical RFT agentic environments for LLMs can integrate both deterministic and stochastic transitions—deterministic for operations like sentence or token concatenation, and stochastic for environment feedback, such as tool execution results or next user queries. Since LLMs generate tokens to form sentences, RFT can operate at both the action-level (sentence-level) and token-level, whereas traditional RL settings are typically action-level. Another major difference is that RL is known for its low sample efficiency, while RFT requires relatively fewer, high-quality samples.

\paragraph{Implementations} Since we know the difference between RFT and RL, we can introduce lots of fine-tuning techniques to better show how RFT works and what is reflected in many existing research. The first is LoRA, which freezes the base model and injects trainable rank decomposition matrices into the model. LoRA is vital for fine-tuning LLMs, especially for LaMAS, because it can optimize the policy efficiently with minimal parameter updates and also makes LLM-based multi-agent training affordable and ideal. \reviewchange{As a single model can have different LoRA adapters for different tasks, multi-agent systems can be made using only one base model equipped with multiple adapters \citep{apple-intelligence-foundation-language-models}.} Secondly, it becomes necessary to implement divergence constraints during training to prevent the policy update from straying too far from the initial or reference policy, which possesses the pretrained abilities. PPO and related objectives encourage conservative policy updates, although strict improvement guarantees require additional assumptions. In these objectives, we can also modify the KL divergence from between $\pi_{\text{old}}$ to between $\pi_\text{ref}$ as GRPO \citep{shao2024deepseekmathpushinglimitsmathematical} does in \eqref{eq:GRPO}. 

\paragraph{Prevailing Objective Functions} To further elaborate on the objective functions used in RFT, we will introduce several prominent methods, including GRPO, DAPO, and VAPO. These objective functions are designed to optimize the policy updates while ensuring stability and efficiency during the fine-tuning process. 

GRPO utilizes the objective function as shown in \eqref{eq:GRPO} \citep{shao2024deepseekmathpushinglimitsmathematical}. As stated in GRPO, $q,o$ are questions and outputs sampled from the question dataset and the old policy $\pi_\text{old}$. This approach dispenses with the need for value function approximation, opting instead to compute a baseline through the average rewards of multiple sampled trajectories. Regularization is achieved by directly incorporating the KL divergence between the trained policy and the reference policy into the loss function.
\begin{small}
\begin{equation}
\begin{aligned}
    J_{GRPO}(\theta) &= \mathbb{E}_{[q \sim P(Q), \{o_i\}_{i=1}^G \sim \pi_{\theta_\text{old}}(O|q)]}\\ &\frac{1}{G} \sum_{i=1}^G \frac{1}{|o_i|} \sum_{t=1}^{|o_i|} \Bigg\{ \min \bigg[r_{i,t}(\theta) \hat{A}_{i,t}, \text{clip} \bigg( r_{i,t}(\theta), 1 \pm \epsilon \bigg) \hat{A}_{i,t} \bigg] - \beta \mathbb{D}_{KL} \big[ \pi_{\theta} \| \pi_\text{ref} \big] \Bigg\},\\ 
\end{aligned}\label{eq:GRPO}
\end{equation}

\begin{equation*}
    \text{where}\ r_{i,t}(\theta) = \frac{\pi_{\theta}(o_{i,t}|q, o_{i,<t})}{\pi_{\theta_{\text{old}}}(o_{i,t}|q, o_{i,<t})},\;\hat{A}_{i,t} = \frac{R_i - \text{mean}(\{R_i\}_{i=1}^G)}{\text{std}(\{R_i\}_{i=1}^G)}.
\end{equation*}
\end{small}

DAPO advances upon GRPO and adopts the objective function as shown in \eqref{eq:DAPO} \citep{yu2025dapoopensourcellmreinforcement}. It promotes exploration by increasing the ceiling clip parameter $\epsilon_\text{high}$. DAPO further enhances its sampling strategy through dynamic sampling, which involves over-sampling and filtering out prompts with accuracy values of 1 and 0, ensuring that all prompts in the batch provide effective gradients. Additionally, DAPO implements a token-level policy gradient loss in the long-CoT RL scenario.
\begin{small}
\begin{equation}
\begin{aligned}
J_{\text{DAPO}}(\theta) &= \mathbb{E}_{(q, a) \sim \mathcal{D}, \{o_i\}_{i=1}^G \sim \pi_{\text{old}}(\cdot | q)} \Bigg[ \frac{1}{\sum_{i=1}^G |o_i|} \sum_{i=1}^G \sum_{t=1}^{|o_i|} \min \Big( r_{i,t}(\theta) \hat{A}_{i,t}, \text{clip} \Big( r_{i,t}(\theta), 1 - \epsilon_{\text{low}}, 1 + \epsilon_{\text{high}} \Big) \hat{A}_{i,t} \Big) \Bigg], \\
&\text{s.t.}\ 0 < \Big| \Big\{ o_i|\text{is\_equivalent}(a, o_i) \Big\} \Big| < G.
\end{aligned}\label{eq:DAPO}
\end{equation}
\end{small}

VAPO extends DAPO and employs the objective function presented in \eqref{eq:VAPO} \citep{yue2025vapoefficientreliablereinforcement}. It further enhances the training process by incorporating an additional negative log-likelihood (NLL) loss for correct outcomes sampled during training. This additional loss term improves the utilization efficiency of positive samples, thereby optimizing the overall learning performance.
\begin{small}
\begin{equation}
\begin{aligned}
L(\theta) = &-\frac{1}{\sum_{i=1}^G |o_i|} \sum_{i=1}^G \sum_{t=1}^{|o_i|} \min \Big( r_{i,t}(\theta) \hat{A}_{i,t}, \text{clip} \Big( r_{i,t}(\theta), 1 - \epsilon_{\text{low}}, 1 + \epsilon_{\text{high}} \Big) \hat{A}_{i,t} \Big) + \mu * L_{\text{NLL}}(\theta),\\
&\text{where}\ L_{\text{NLL}}(\theta)=- \frac{\sum_{o_i \in \mathcal{T}} \sum_{t=1}^{|o_i|} \log \pi_{\theta}(a_t | s_t)}{\sum_{o_i \in \mathcal{T}} |o_i|}.
\end{aligned}\label{eq:VAPO}
\end{equation}
\end{small}

\subsection{LLM-based Multi-Agent Systems (LaMAS)}
\label{sec:review:lamas}

The concept of LaMAS has recently gained prominence \citep{yang2024llmbasedmultiagentsystemstechniques}, with numerous applications and inference frameworks emerging to address complex agentic problems \citep{2503.02068}. Unlike traditional MARL, where agents act synchronously with uniform decision weights, LaMAS introduces hierarchical organization and asynchronous execution. Agents in LaMAS can dynamically decompose tasks, adapt workflows based on execution dependencies, and coordinate actions in a decentralized yet goal-aligned manner.

Existing optimization methods of LaMAS can be categorized into two types as below:
\begin{itemize}
    \item \textbf{Tuning-free techniques.} A prevalent optimization strategy for LaMAS involves tuning-free methods that do not update agent parameters. Examples include prompt engineering \citep{fernando2023promptbreederselfreferentialselfimprovementprompt}, in-context learning \citep{NEURIPS2024_5d1f0213}, and self-evolution \citep{yang2025agentnetdecentralizedevolutionarycoordination,huang2024agentcodermultiagentbasedcodegeneration}. These techniques rely on iterative refinements of agent profiles or prompts to enhance task-specific performance.
    \item \textbf{Parameter fine-tuning.} A more sophisticated approach involves updating agent parameters or network configurations to optimize performance. This paradigm integrates deep learning and specific methodologies, such as multi-agent debating to generate high-quality training datasets \citep{subramaniam2025multiagentfinetuningselfimprovement} or fine-tuning programming modules for improved workflow organization \citep{khattab2024dspy}. CORY \citep{NEURIPS2024_1c2b1c8f} duplicates the LLM into two agents, i.e., a pioneer and an observer, that interact through knowledge transfer and role exchange, improving training stability and performance. However, many parameter fine-tuning methods lack a theoretical foundation in RL, relying instead on ad-hoc techniques. SiriuS~\citep{zhao2025siriusselfimprovingmultiagentsystems}, a self-improving, multi-agent optimization framework using bootstrapped reasoning, iteratively refines agent behaviors by leveraging successful reasoning trajectories and augmenting unsuccessful ones. MARTI~\citep{marti2025} develops an RL-oriented framework for multi-agent LLM training and inference, but still treats agents largely through separate policy trainers and full-parameter tuning, which can be costly for scalable LaMAS optimization. The most related work, MAPoRL~\citep{park2025maporlmultiagentpostcotrainingcollaborative}, rooted in MARL, pioneers in applying multi-agent PPO to post-train LLMs, but does not consider the unique characteristics of LaMAS and assumes homogeneous agent behavior through collaborative debate, diverging from real-world scenarios where agents often perform diverse or orthogonal roles to solve complex problems. This article will focus on RFT and MARL-related methodologies in subsequent sections, providing a principled paradigm for general LaMAS reinforcement.
\end{itemize}

\subsection{Multi-Agent Reinforcement Learning (MARL)}
\label{sec:review:marl}

\subsubsection{Decentralized Partially Observable Markov Decision Process} 
\label{sec:review:marl:dec-pomdp}
MARL is a variant of RL, which includes multiple agents acting \textbf{simultaneously} once receiving the environment observations, and it's often modeled by a certain form of MDP -- \textit{Decentralized POMDP (DEC-POMDP)} \citep{10.5555/2967142} $\langle\mathcal{N},{\mathcal{S}},\boldsymbol{\mathcal{O}},\boldsymbol{\mathcal{A}},\mathcal{T},\mathcal{R},\gamma\rangle$, where $\mathcal{N} = \{1, \dots, n\}$ is the set of agents \textbf{without any order constraint}, ${\mathcal{S}}$ is the state space, $\boldsymbol{\mathcal{O}} = \prod^n_{i=1}\mathcal{O}^{i}$ is the joint observation space, which is a product of local observation spaces at global state $s$, $\boldsymbol{\mathcal{A}} = \prod^n_{i=1}\mathcal{A}^{i}$ is the joint action space, which is similarly a product of local action spaces, and for the others, we can borrow the explanation from Section \ref{sec:review:rft:language-augmented-mdp}.

\subsubsection{Main Approaches of MARL}
\label{sec:review:marl:approaches}

\paragraph{Independent Learning} One of the most intuitive approaches to implementing MARL is through independent learning. In this setting, each agent operates as an independent entity, possessing its own policy and value function. Other agents are treated as part of the environment, and there is no direct communication between agents. Consequently, each agent optimizes its actions based solely on its own observations and rewards, without considering the policies of other agents.

This approach allows for the direct application of single-agent reinforcement learning methods to each agent. Examples include independent Q-learning \citep{10.5555/3091529.3091572,10.5555/1483085}, independent policy gradient \citep{NEURIPS2020_3b2acfe2}, and independent AC methods. These methods are straightforward to implement and can be easily scaled to various domains. However, independent learning has significant drawbacks. It can lead to instability and difficulty in converging to an optimal solution. For instance, some agents may develop selfish policies or become "lazy" in collaborative tasks, resulting in what is known as a social dilemma.

Early experiments have demonstrated that, for certain tasks, cooperative agents outperform independent agents, especially when communication is absent \citep{10.5555/3091529.3091572}. The cost of communication and the optimization of information exchange strategies require careful consideration in collaborative learning. Despite the challenges of instability and poor convergence, independent learning remains a fundamental and viable approach to MARL.

\paragraph{Centralized Training with Decentralized Execution} Apart from independent learning, another way to MARL is by centralized training. During training, agents share a joint global value function $Q^{\bm{\pi}}(\mathbf{o}, a^{1}, \dots, a^{n})$ which takes into account the environment state and actions of other agents. For its design, $Q$ can either output the value for the joint action and marginalize out for a specific action, as in COMA \citep{Foerster_Farquhar_Afouras_Nardelli_Whiteson_2018}, or directly output the value for action $a^i$ by agent $i$, as in MADDPG \citep{lowe2017multi}. The latter approach allows for arbitrary reward structures for separate agents. Once the value learning component is established, policy optimization can be performed using the policy gradient: $\nabla_{\theta^i} J(\theta^i) = \mathbb{E}_{s \sim p^{\bm{\mu}}, a^i \sim \bm{\pi}^i} \left[ \nabla_{\theta^i} \log \bm{\pi}^i(a^i|o^i) Q^{\bm{\pi}}(\mathbf{o}, a^{1}, \dots, a^{n}) \right]$ to update the policies.

As mentioned in Section \ref{sec:review:rft:rl}, PPO has consistently been one of the most effective RL methods. Fortunately, it also extends well to multi-agent scenarios, leading to Multi-Agent PPO (MAPPO) \citep{yu2022the}. 
Then, the objective of MAPPO is derived as
\begin{small}
\begin{gather}
    \label{eq:mappo-objective}
        \maximize_{\theta} \sum_{i=1}^n \mathbb{E}_{s\sim \rho_{\bm{\pi}_{\theta_\text{old}}}, \bm{a}\sim\bm{\pi}_{\theta_\text{old}}} \left[\min \left( \frac{\pi^{i}_{\theta}(a^{i}|s)}{\pi^{i}_{\theta_\mathrm{old}}(a^{i}|s)} A^{\bm{\pi}_{\theta_\mathrm{old}}}(s, \bm{a}),\text{clip} \left( \frac{\pi^{i}_{\theta}(a^{i}|s)}{\pi^{i}_{\theta_\mathrm{old}}(a^{i}|s)}, 1 \pm \epsilon \right) A^{\bm{\pi}_{\theta_\mathrm{old}}}(s, \bm{a}) \right)\right].
\end{gather}
\end{small}

However, this approach has a strong assumption that all agents share parameters, which leads to identical action spaces among agents. In the context of LLMs, it is common for different LLMs to have distinct action spaces. For instance, different LLMs may possess unique vocabularies, and some LLMs might even have different input-output formats, such as LLMs designed for ranking tasks. Moreover, MAPPO does not provide a rigorous guarantee of monotonic improvement, as a local improvement in one agent's policy can potentially degrade the overall team reward. HATRPO and HAPPO addressed this issue by sequentially updating the policies \citep{kuba2022trust}.

\paragraph{Communication and Coordination} Another intriguing approach to MARL involves directly learning the communication among agents. For instance, designing a communication module or gate module can help establish an effective explicit communication and coordination mechanism within the multi-agent system.

One seminal work in this area is CommNet \citep{NIPS2016_55b1927f}, which introduced a novel trainable controller $\bm\Phi$ that maps global state-views s to joint policies a. In this mapping, each agent has access to the hidden states of other agents, thereby influencing its own policy. This approach can be seamlessly integrated with RL algorithms and ideally results in an optimal communication module.

In the context of LLM-based multi-agent systems, such communication modules can take the form of an "orchestra agent". This agent is specifically designed and trained to act as a task and information router, facilitating effective collaboration among the agents. Currently, a popular solution involves designing a well-structured multi-agent framework that includes a planner or orchestration agent to enhance collaboration. However, most of these attempts focus on improving performance during inference without incorporating further training.

\paragraph{Sequential Modeling} Another intriguing perspective on understanding MARL is through sequential modeling (SM). The connection between MARL and SM is grounded in the Multi-Agent Advantage Decomposition Theorem \citep{NEURIPS2021_6fe6a8a6} as below:
\paragraph{Theorem 1}(Multi-Agent Advantage Decomposition Theorem \citep{NEURIPS2021_6fe6a8a6}). \textit{For any predefined permutation of $n$ agents, for any state $s \in \mathcal{S}$ and joint action $\bm{a} \in \boldsymbol{\mathcal{A}}$, the following always holds:}
\begin{align}
    \label{eq:advantage-decomposition}
    A^{\bm\pi}(s,\bm{a}^{1:n}) = \sum_{m=1}^n A^{\bm\pi}(s,\bm{a}^{1:m-1},\bm{a}^{m}).
\end{align}
This theorem reveals that if an agent is aware of the actions taken by its predecessors, maximizing the agent's local advantage is equivalent to maximizing the joint advantage. Building on this insight, a new multi-agent sequential decision-making paradigm has been proposed, exemplified by the Multi-Agent Transformer (MAT) \citep{MAT}. Within this paradigm, the approach to approximating the value function remains consistent with conventional MARL methods, specifically by minimizing the empirical Bellman error below\footnote{In MAT, the \textit{hat}~$\hat{\cdot}$~means some encoding of the variable.} as 
\begin{align}
    \label{eq:mat-value-objective}
    L(\phi) = \frac{1}{nT} \sum_{i=1}^{n} \sum_{t=0}^{T-1} \left[ r_t(\bm{o}_t, \bm{a}_t) + \gamma V_{\bar{\phi}}(\hat{\bm{o}}_{t+1}^{i}) - V_{\phi}(\hat{\bm{o}}_{t}^{i}) \right]^2,
\end{align}
where $\bar\phi$ is the parameters of the target network, which are non-differentiable and updated at intervals. But for the policy optimization, the modified clipping PPO objective goes as 
\begin{small}
\begin{gather}
    \label{eq:mat-policy-objective}
    \hspace{-5pt}
    L(\theta) = \frac{1}{nT} \sum_{i=1}^{n} \sum_{t=0}^{T-1} \min \left[ \bm{r}_t^{i}(\theta) \hat{A}_t, \text{clip}\left(\bm{r}_t^{i}(\theta), 1 \pm \epsilon \right)\hat{A}_t\right], \text{where} \ \bm{r}_t^{i}(\theta) = \frac{\pi_{\theta}^{i} \left( a_t^{i}|\hat{\bm{o}}_t^{1:n}, \hat{\bm{a}}_t^{1:i-1} \right)}{\pi_{\theta_{\text{old}}}^{i} \left( a_t^{i}|\hat{\bm{o}}_t^{1:n}, \hat{\bm{a}}_t^{1:i-1} \right)}.
\end{gather}
\end{small}

An intuitive way to understand this paradigm is to consider that, before acting \textbf{simultaneously}, agents make decisions (not take actions) sequentially in an arbitrary order or permutation, based on what their predecessors have planned or decided to do. They then execute their actions accordingly. For instance, in a football game, it is akin to players stating their intended actions one by one during halftime. Once on the field, they follow through with the sequential decisions they have just made.

\paragraph{Networked Agents} One more novel cooperative MARL paradigm involves networked agents, which operate in fully decentralized settings. In this framework, agents perceive local environmental observations and synchronize information locally with neighboring agents over a network \citep{zhang2018fullydecentralizedmultiagentreinforcement}. A consensus mechanism has been introduced to the paradigm, expanding its applicability to real-world domains with strict privacy requirements \citep{chen2022communicationefficient,varela2025networked}. By enabling localized information sharing and coordination, networked agents enhance system robustness and adaptability while preserving data privacy, making this paradigm particularly promising for applications in decentralized and privacy-sensitive environments, especially for LaMAS in use.

\section{MARFT Overview}
\label{sec:overview}

Based on the review of RFT, LaMAS, and MARL, in this section, we will extend this foundation to the context of MARFT, highlighting the problem statement, unique considerations, and additional complexities that arise in this domain.

\subsection{Problem Statement and Flex-MG}
\label{sec:overview:problem-statement}

MARFT integrates the three key principles comprehensively reviewed in Section \ref{sec:review}, adapting them to the context of RFT for LaMAS. In MARFT, agents operate within a framework that allows asynchronous yet coordinated execution, reflecting the dependencies inherent in agentic problem-solving (as detailed in Section \ref{sec:review:lamas}). This design is particularly suitable for addressing complex tasks where execution dependencies necessitate dynamic adaptation and partial observability. Thus, the algorithms developed for MARFT serve dual objectives: optimizing concurrent actions and learning systemic organization and dynamic task decomposition workflow.

\begin{figure}[htb]
    \centering
    \includegraphics[width=0.99\linewidth]{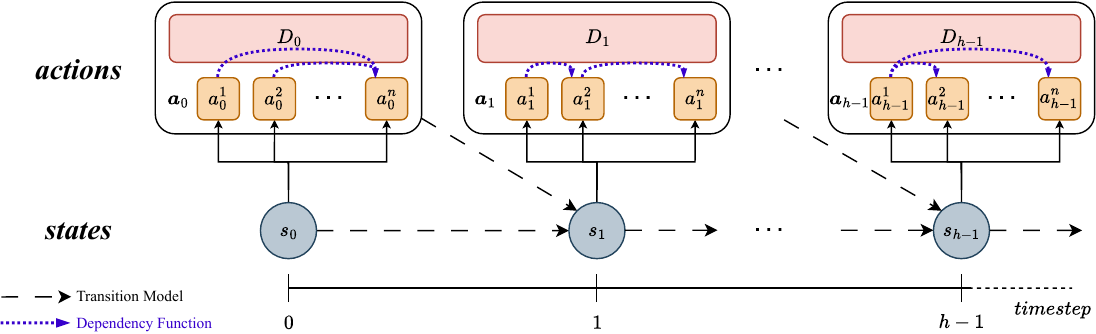}
    \caption{A detailed illustration of the dynamics of a Flex-MG. The dependency function (dashed purple line) can vary across timesteps.}
    \label{fig:flex-mg}
\end{figure}

To enhance problem formulation and accommodate MARFT, we propose \textit{Flexible Markov Game (Flex-MG)}, illustrated in Figure~\ref{fig:flex-mg} and denoted as $\langle\mathcal{V}, \mathcal{N},{\mathcal{S}},\boldsymbol{\mathcal{A}},\mathcal{T},\mathcal{R},\gamma,\mathcal{D}\rangle$. Here, $\mathcal{V}$ is the vocabulary, $\mathcal{N}=\{1,...,n\}$ is the agent composition of a LaMAS \textbf{with some organization}, i.e., some order constraint among agents, $\mathcal{S}$ is the state space, and $\boldsymbol{\mathcal{A}} = \prod^n_{i=1}\mathcal{A}^{i}$ is the joint action space. The transition function $\mathcal{T} \colon \mathcal{S} \times \boldsymbol{\mathcal{A}} \rightarrow \mathcal{S}^\prime$, written as $P(s^\prime|s, \bm{a})$, models state changes given joint actions. The reward function $\mathcal{R} \colon \mathcal{S} \times \boldsymbol{\mathcal{A}} \rightarrow \mathbb{R}$, denoted $R(s, \bm{a})$, assigns rewards based on states and joint actions. The discount factor $\gamma$ accounts for temporal reward decay.

Flex-MG is distinguished by its introduction of a \textbf{dependency function} $\mathcal{D} \colon \boldsymbol{\mathcal{A}} \times \boldsymbol{\mathcal{A}} \rightarrow \{0,1\}$, which explicitly models inter-agent dependencies. Specifically, $\mathcal{D}(a^i, a^j) = 1$ indicates that the action of agent $j$ is conditionally dependent on the action of agent $i$. As a result, the decision-making process of agent $j$ is influenced not only by the global state $s$ but also by the actions of all agents $k$ such that $k \in \{l|\mathcal{D}(a^l, a^j) = 1\}$. These dependencies can be integrated into agent $j$'s input via concatenation or other fusion methods, thereby forming an enriched input for the agent's policy.
Crucially, the dependency function $\mathcal{D}$ can vary across timesteps, potentially governed by an orchestration agent or a dynamic routing mechanism. This flexibility captures the evolving coordination structures in LaMAS and embodies the "flexible" nature of the proposed framework. Notably, when $\mathcal{D}(a^i, a^j) = 0$ for all $i, j$, Flex-MG reduces to a standard multi-agent decentralized setting where agents act independently.

\subsection{Categorizations of LaMAS in MARFT}
\label{sec:overview:lamas-categorizations}

In the context of MARFT, the unique characteristics of LaMAS necessitate a detailed categorization. We classify LaMAS based on three key dimensions: parameter sharing, execution synchronicity, and update scheme. This categorization not only captures the distinct features of LaMAS but also provides a clear direction for the development of MARFT algorithms.

\paragraph{Parameter sharing} Parameter sharing is a critical consideration in multi-agent systems, particularly in traditional MARL. However, in LaMAS, parameter differentiation strategies introduce additional complexity. Agents may either share parameters or maintain distinct configurations, each with unique implications for optimization.

For parameter sharing, agents share an identical base model, which can be fully fine-tuned or LoRA fine-tuned on a single adapter for task-specific roles. For non-parameter sharing, agents may use different base models with fine-tuning (full or LoRA) or share a common base model with role-specific adapter modules. While parameter sharing streamlines optimization and alleviates memory pressure, it can introduce challenges such as opposite optimization directions or parameter drift, where conflicting updates degrade performance. Despite these risks, parameter sharing remains essential for efficient LaMAS fine-tuning, particularly when combined with techniques like LoRA, which isolate task-specific adaptations from the base model. 

\paragraph{Execution synchronicity} Execution synchronicity represents another critical dimension for classifying LaMAS. Unlike traditional cooperative games in multi-agent systems, where agents typically act synchronously, LaMAS often operates under asynchronous or hybrid execution paradigms. This distinction arises from the inherent dependencies in agentic tasks, where execution outcomes frequently rely on partial or sequential results.

In synchronous LaMAS, agents execute actions concurrently, assuming full observability and independence of execution results, while in asynchronous LaMAS, agents act independently, with execution triggered by task-specific dependencies or environmental signals. The latter paradigm is prevalent in real-world applications, as it accommodates dynamic workflows and partial observability. Hybrid LaMAS combines synchronous and asynchronous elements to balance efficiency and flexibility.

\paragraph{Update simultaneity} In addition to execution synchronicity during inference, the choice of update scheme during training further distinguishes LaMAS configurations. Specifically, LaMAS can be categorized into two types based on how agent updates are orchestrated.

In a LaMAS with synchronous update manner, all agents are updated concurrently. Though simple and easy to implement, it introduces non-stationarity into the optimization process and suffers from severe off-policy issues as later agents may base their updates on outdated policies of earlier agents. Each agent's policy updates are based on rapidly changing system dynamics, complicating convergence and potentially leading to oscillations in performance. On the other hand, in sequential (i.e., agent-by-agent) update LaMAS, agents are updated sequentially, with each agent's update based on the most current policies of others. This method can reduce non-stationarity and off-policy effects; monotonic-improvement results require the assumptions of the relevant trust-region analysis and do not automatically extend to finite-sample clipped PPO. However, it may increase computational overhead due to the sequential nature of updates.

\subsection{Differences and Resulting Challenges}
\label{sec:overview:differences-and-challenges}

Due to the distinct characteristics of LaMAS compared to traditional multi-agent systems in MARL, implementing large-scale LaMAS within a MARL framework presents significant challenges, and classic MARL methods often fail to effectively enhance the performance of LaMAS.\footnote{In fact, there is very little research about implementing classic MARL methods, such as COMA, MADDPG, MAPPO, HAPPO, etc., to LaMAS.} Below, we list the key differences between MARFT and traditional MARL in table \ref{tab:marft-differences}, and expand the discussion in detail, which hopefully can guide the methodology and implementation in the following section. 

\begin{table}[htbp]
    \begin{center}
    \caption{Differences between MARFT and traditional MARL.}
    \label{tab:marft-differences}
        \begin{tabular}{lll}
            \toprule 
            \multicolumn{1}{l}{}  &\multicolumn{1}{l}{\bf MARL}  &\multicolumn{1}{l}{\bf MARFT} \\
            \midrule
            Execution Asynchronicity &Synchronous &Asynchronous \\
            Utilities &Task-specific only &Agentic with original language ability \\
            Agent Identity &Unspecified, usually id &Specified system prompts as profiles \\
            Heterogeneity &Parameters &In-out formats, externals, parameters \\
            System Organization &Static &Static/Dynamic\\
            Optimization Space &Small &Large and variable \\
            \bottomrule
        \end{tabular}
    \end{center}
\end{table}

\paragraph{Execution Asynchronicity} One of the most significant differences between conventional MARL and MARFT is the nature of agent action executions. In traditional cooperative MARL, agents' actions are typically executed simultaneously. However, in MARFT, they are executed asynchronously. In some cases, the actions of certain agents may even depend on the outcomes of other agents' actions. For example, in a collaborative coding assistant system, one LLM agent generates a function prototype, and another agent asynchronously refines it based on the first agent's output, demonstrating both asynchronicity and result dependency. As a result, conventional MARL methods, which assume synchronous actions, may not be directly applicable or effective in LLM-based multi-agent systems.

\paragraph{Utilities} Unlike agents in typical MARL problems, LLMs are initially designed for language processing rather than specific agentic tasks. However, the value- or reward-guided nature of RL means that when using value-based RL algorithms to optimize or extract policies, the derived optimal policy often prioritizes actions that maximize the value function. This can lead to policies that generate high-reward actions or tokens but are incomprehensible to humans. Though some attempts to extract a policy together with entropy regularization have been made to improve its agentic intelligence without harming the text capability \citep{wen2024entropyregularizedtokenlevelpolicyoptimization}, it is still a point that requires additional attention when we try to implement value-based methods to optimize LLMs.

\paragraph{Characteristic Profiles} The transition from single-agent to multi-agent systems in typical MARL benchmarks, such as Multi-Agent MuJoCo, often involves splitting a single agent into multiple components without additional modifications. In contrast, LLMs require a more nuanced approach. When decomposing tasks into orthogonal sub-tasks, each LLM agent needs a profile to define its role and capabilities, enabling it to generate actions consistent with its assigned character. This profile can be human-specified or learned by the agent through natural evolution. Consequently, the joint observation space in MARFT is augmented with profiles, taking the form \texttt{[agent profile + env observation (+ agent-specific guidance)]}. This modification is crucial when designing a MARL training framework.

\paragraph{Heterogeneity} In conventional MARL benchmarks, heterogeneity is typically characterized by differences in agent structures and non-parameter-sharing schemes. However, LLM-based systems introduce a higher level of complexity. First, LLMs themselves can vary significantly in terms of model structures, parameters, input and output formats (\eg, LLMs vs. Vision-Language Models), and vocabularies. Second, agents may have access to different external tools or devices, reflecting real-world scenarios. For instance, in designing a multi-agent system for a mobile operating system, some agents may have access to both local and remote search engines, while others may rely on proprietary databases or tools from other companies. This heterogeneity complicates training, making it more difficult and unstable. 

\paragraph{System Organization} Traditional MARL tasks are often set in static simulated environments. In contrast, LLM-based multi-agent systems are designed for real-life agentic tasks with higher uncertainty. These tasks can be decomposed in various ways, leading to different multi-agent systems with distinct populations and roles. Moreover, agents may exhibit sequential dependencies and contextual relationships when solving sub-tasks. For example, one agent's action may depend on the outcome of another agent's action. This dynamic organization can be either human-designed or agent-explored (\eg, through training), adding another layer of complexity to the design process.

\paragraph{Optimization Space} If we take one LLM generation as an action, the observation space is exponentially vast as $o \in \mathcal{O} \subseteq \mathcal{V}^L$, whose complexity is $O(\lvert\mathcal{V}\rvert^L)$, where $\mathcal{V}$ is the vocabulary of the acting LLM and $N$ is the token length of the generated action, and both of them vary with different acting LLMs. It makes the value function hard to converge with stability. If we take one token as an action, the complexity can be reduced to $O(\lvert\mathcal{V}\rvert\times L)$, but it will become a task with super sparse reward outcome so that the value function should also be meticulously learned, as $L$ can be extremely large. This is also reflected in VAPO and is mitigated by value pretraining \citep{yue2025vapoefficientreliablereinforcement}. 
Fortunately, not all LLM-related tasks face such large optimization spaces. For instance, in embodied AI with LLMs, predefined actions (\eg, "go to the kitchen," "grab a coffee") are often provided, significantly narrowing the action space \citep{pmlr-v202-carta23a,tan2024true}.

These differences collectively make the optimization of LaMAS more complex than traditional MARL problems. MARFT is designed to tackle the difficulties. To generally optimize the LaMAS regardless of the highly dynamic workflow or organization, MARFT reframes the problem as a sequential decision-making problem despite the dynamic organization. It preserves pretrained utilities by applying clipping, preventing excessive policy drift. Agent profiles are encoded to activate specific capabilities, and no modality assumptions are made, as all agent actions are expressed in tokens from their respective vocabularies.

\section{Multi-Agent Reinforcement Fine-Tuning}
\label{sec:marft}

Building on the formulations outlined in the previous section, we now turn to the core methodologies that enable MARFT to address the unique challenges of optimizing LaMAS. 

\subsection{MARFT Instantiations}
\label{sec:marft-methods:methodology}

\begin{figure}[htb]
    \centering
    \includegraphics[width=0.99\linewidth]{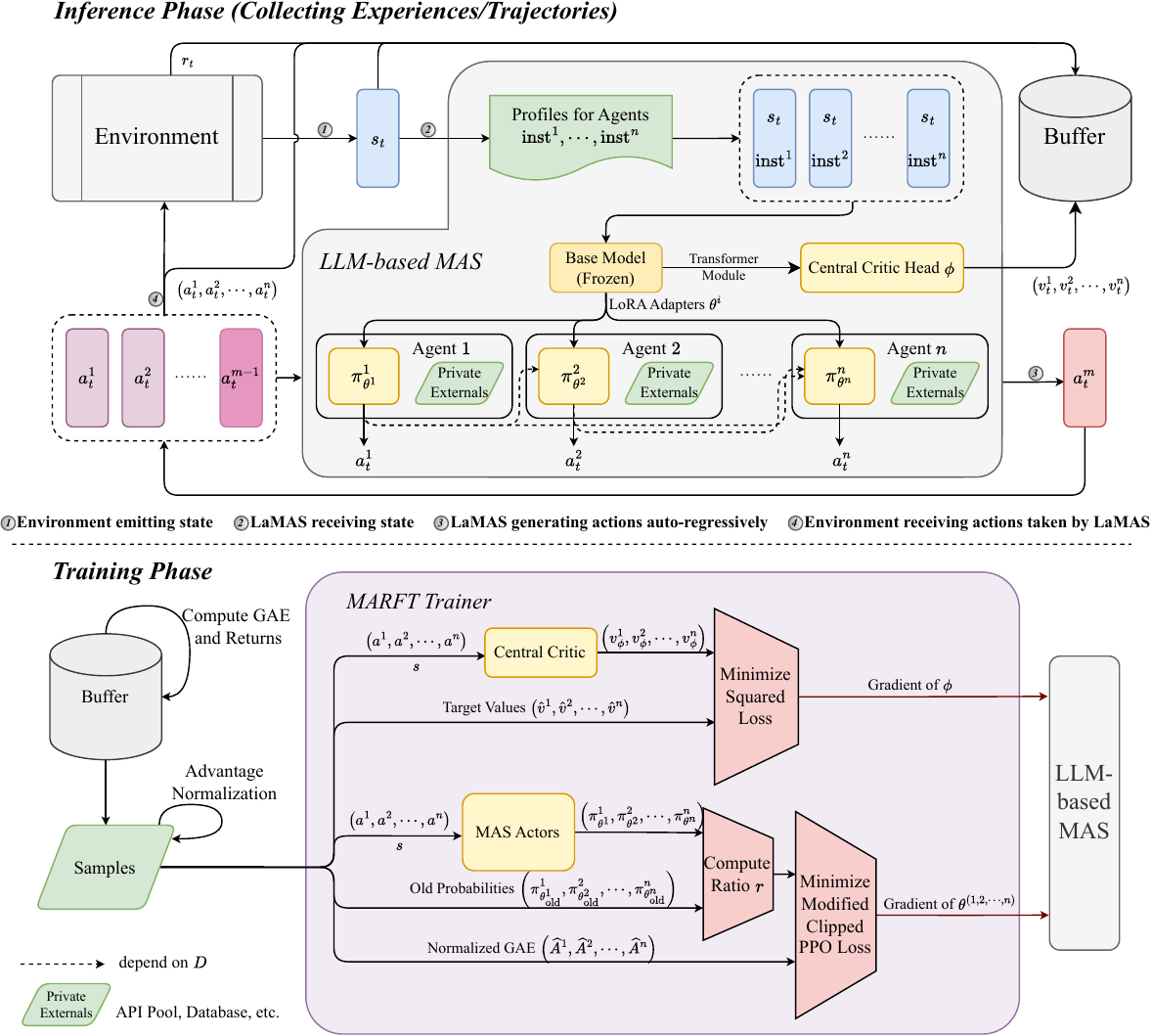}
    \caption{The procedure of Multi-Agent Reinforcement Fine-Tuning. Inference and training are conducted in an alternating manner. Each agent within the MAS can have its own private APIs, tool pool, database, and other resources.}
    \label{fig:marft-process}
\end{figure}

Inheriting the SM-style re-modeling via the multi-agent advantage decomposition theorem introduced in the last section, MARFT follows the procedure demonstrated in Figure~\ref{fig:marft-process}. 
Given any organizational or task-solving workflow of a LaMAS, theorem \eqref{eq:advantage-decomposition} allows MARFT to reframe it as a sequential decision-making process. When collecting interactive trajectories, the LaMAS generates actions with an auto-regressive problem-solving style. The "encoder" constructs local roll-out states using agent profiles and other relevant information, while the "decoder" combines these states with dependent predecessors' actions to generate the current agent's action $a^m \sim p_{\pi^m}(a^m|s,\bm{a}^{1:m-1})$. 

To instantiate the framework, we first present MARFT on the action level, a variant designed for targeted, action-level policy optimization in LaMAS. During training, MARFT adopts an optimization objective in \eqref{eq:marft-objective}\footnote{Our current MARFT instantiation employs a sequential decision-making process for scalability. Extending it to handle the parallel rollouts permitted by Flex-MG (i.e., a general DAG structure) would necessitate a more complex optimization objective to tackle off-policy problems, which we leave as a direction for future work.}. The predecessor-conditioned objective is motivated by the advantage-decomposition and trust-region analyses underlying MAT and HAPPO. In the practical finite-sample implementation, clipped PPO encourages conservative updates but does not provide an unconditional monotonic-improvement guarantee. 
\begin{align}
    &L(\bm\theta) = \frac{1}{nT} \sum_{i=1}^{n} \sum_{t=0}^{T-1} \min \left[ \bm{r}_t^{m}(\bm\theta) \hat{A}_t, \text{clip}\left(\bm{r}_t^{m}(\bm\theta), 1 \pm \epsilon \right)\hat{A}_t\right], \label{eq:marft-objective}\\ 
    &\text{where} \ \bm{r}_t^{m}(\bm\theta) = \frac{\pi_{\theta^m}^{m} \left( a_t^{m}|s_t, \hat{\bm{a}}_t^{1:m-1} \right)}{\pi_{\theta^m_{\text{old}}}^{m} \left( a_t^{m}|s_t, \hat{\bm{a}}_t^{1:m-1} \right)}. \nonumber
\end{align}

\textbf{Action Normalization.} As shown in equation \eqref{eq:llm-action-probability}, the probability of each token $P\left(w|\cdot \right)$ is less than (or equal to) 1. As a result, longer action tends to have a lower joint probability, even though it is more reasonable than other actions in the task. \reviewchange{To address the issue, token normalization and word normalization divide the joint action probability in \eqref{eq:llm-action-probability} by a constant related to its token or word length \citep{tan2024true}. This gets rid of the strong assumption that all the actions have similar lengths, which in LLMs can be an evil condition, because in agentic tasks, it is all too common that solutions generated can be either super long with a chain-of-thought or super short with a direct answer.}
\begin{align}
    \text{log}P_{\pi^i}\left(a^i|s,\bm{a}^{\mathcal{N}(i)}\right) &= \text{log}P_{\pi^{i}}\left(w^{i,1}, \ldots, w^{i,L^i}|s,\bm{a}^{\mathcal{N}(i)}\right) \nonumber\\
    &= \sum_{j=1}^{L^i} \text{log}P_{\pi^i}\left(w^{i,j} | s,\bm{a}^{\mathcal{N}(i)}, w^{i,1}, \ldots, w^{i,j-1}\right). \label{eq:llm-action-probability}
\end{align}
\textbf{Agent-by-agent Update.} Updating every policy during each training epoch may introduce implicit off-policy effects as predecessor policies change. This issue is especially relevant in LaMAS, where agents depend on other agents' actions. Drawing on A2PO \citep{wang2023order}, agent-by-agent updates can reduce this mismatch and motivate more stable optimization. Their theoretical improvement properties depend on the assumptions of the corresponding analysis; finite-sample clipped PPO does not automatically inherit a strict guarantee. 

\textbf{Trajectory-level Optimization.} For language models, the definition of action from the RL perspective can be dynamic. Direct trajectory-level optimization, such as GRPO with outcome supervision \citep{shao2024deepseekmathpushinglimitsmathematical} and its variant DAPO \citep{yu2025dapoopensourcellmreinforcement}, can be viewed as an action-level optimization with non-zero reward only at the end of the trajectory in the MDP or a normal action-level optimization in a single-step variant of the MDP. Consequently, trajectory-level optimization can be seamlessly integrated into MARFT by following the same idea.

\begingroup
\tokenrevision
On top of assigning credit to each agent action, we obtain a
finer-grained instantiation by treating every generated token as an
action. Let $z_t^{m,j}\triangleq
(s_t,\bm a_t^{1:m-1},w_t^{m,1:j-1})$ denote the history immediately
before agent $m$ emits token $j$, and let
$\bar z_t^{m,j}\triangleq(z_t^{m,j},w_t^{m,j})$ denote the corresponding
post-token history, where $w_t^{m,1:0}=\emptyset$. We use $\gamma_a=\gamma$ for the Flex-MG environment discount and
$\gamma_w$ for the discount applied only to within-action token transitions,
with $0\leq\gamma_a,\gamma_w\leq1$. Let
$\widetilde G_{\gamma_w}(\tau\mid h)$ denote the return induced by this
two-scale discount schedule from token history $h$ along trajectory $\tau$.
We retain $\widetilde V$ and $\widetilde Q$ to distinguish token-level
conditional values from their action-level counterparts. Starting before a
token is emitted introduces no additional discount relative to the history
immediately after that sampled token; the discount between consecutive
post-token histories is accounted for by the backups below:

\begin{equation}
\begin{aligned}
\widetilde V^{\bm\pi}(z_t^{m,j})
&\triangleq
\mathbb E_{\tau\sim p_{\bm\pi}(\tau\mid z_t^{m,j})}
\left[
\widetilde G_{\gamma_w}(\tau\mid z_t^{m,j})\mid z_t^{m,j}
\right],\\
\widetilde Q^{\bm\pi}(z_t^{m,j},w_t^{m,j})
&\triangleq
\mathbb E_{\tau\sim p_{\bm\pi}
(\tau\mid z_t^{m,j},w_t^{m,j})}
\left[
\widetilde G_{\gamma_w}(\tau\mid \bar z_t^{m,j})
\mid z_t^{m,j},w_t^{m,j}
\right].
\end{aligned}
\label{eq:token-ma-value-functions}
\end{equation}

The pre-token value averages over the next token under the same policy:
\begin{equation*}
\widetilde V_{\gamma_w}^{\bm\pi}(z_t^{m,j})
=\mathbb E_{w\sim\pi^m(\cdot\mid z_t^{m,j})}
\left[\widetilde Q_{\gamma_w}^{\bm\pi}(z_t^{m,j},w)\right].
\end{equation*}
For a realized token, the corresponding post-token value
satisfies $\widetilde V^{\bm\pi}(\bar z_t^{m,j})
=\widetilde Q^{\bm\pi}(z_t^{m,j},w_t^{m,j})$. Intermediate token rewards are
zero, $\gamma_w$ discounts
transitions within an agent action, and $\gamma_a$ discounts environment
transitions. Following the token-by-token expansion, the state-value backup
starts from the post-token history $\bar z_t^{m,j}$:
\begin{equation}
\label{eq:ma-v-bellman-backups}
\widetilde V^{\bm\pi}(\bar z_t^{m,j})
=
\begin{cases}
0+\gamma_w\mathbb E_{w_t^{m,j+1}\sim\pi^m}
 [\widetilde V^{\bm\pi}(\bar z_t^{m,j+1})],
& j<|a_t^m|,\\
0+\mathbb E_{w_t^{m+1,1}\sim\pi^{m+1}}
 [\widetilde V^{\bm\pi}(\bar z_t^{m+1,1})],
& j=|a_t^m|,\ m<n,\\
R_t+\gamma_a\mathbb E_{s_{t+1}\sim P}
 [\widetilde V^{\bm\pi}(z_{t+1}^{1,1})],
& j=|a_t^m|,\ m=n.
\end{cases}
\end{equation}
The state--token-value backup recurses directly through the
next-token value:
\begin{equation*}
\begin{aligned}
&\widetilde Q^{\bm\pi}(z_t^{m,j},w_t^{m,j})\\[-0.25ex]
&\quad=
\begin{cases}
0+\gamma_w\mathbb E_{w_t^{m,j+1}\sim\pi^m}
 [\widetilde Q^{\bm\pi}(z_t^{m,j+1},w_t^{m,j+1})],
& j<|a_t^m|,\\
0+\mathbb E_{w_t^{m+1,1}\sim\pi^{m+1}}
 [\widetilde Q^{\bm\pi}(z_t^{m+1,1},w_t^{m+1,1})],
& j=|a_t^m|,\ m<n,\\
R_t+\gamma_a\mathbb E_{\substack{s_{t+1}\sim P\\
 w_{t+1}^{1,1}\sim\pi^1}}
 [\widetilde Q^{\bm\pi}(z_{t+1}^{1,1},w_{t+1}^{1,1})],
& j=|a_t^m|,\ m=n.
\end{cases}
\end{aligned}
\end{equation*}
Each token expectation is conditioned on its displayed
pre-token history, and the final state expectation uses
$P(\cdot\mid s_t,\bm a_t)$. The optimal-control backups replace the token
expectations by $\max_w$ and $\bm\pi$ by $\star$, while retaining the
environment-transition expectation. During training, realized tokens and
$s_{t+1}$ provide single-sample TD targets for these conditional backups.

For a fixed complete trajectory, the two-scale discount schedule yields
the following within-action discount counts:
\begin{equation}
\begin{aligned}
E_{1,t}^{m,j}(\bm a_t)
&=\sum_{k=m}^{n}|a_t^k|-j-(n-m),\\
E_{2,t}^{m,j}(\bm a_t,\bm a_{t+1})
&=E_{1,t}^{m,j}(\bm a_t)
+\sum_{k=1}^{n}(|a_{t+1}^k|-1).
\end{aligned}
\label{eq:token-discount-exponents}
\end{equation}
For a fixed complete trajectory $\tau$ with an absolutely convergent
action-level return, let
\begin{equation*}
\begin{aligned}
D_{\ell,t}^{m,j}(\tau)
&\triangleq E_{1,t}^{m,j}(\bm a_t)
+\sum_{r=1}^{\ell}\sum_{k=1}^{n}(|a_{t+r}^{k}|-1),\\
\widetilde G_{\gamma_w}(\tau\mid\bar z_t^{m,j})
&=\sum_{\ell=0}^{\infty}\gamma_a^{\ell}
\gamma_w^{D_{\ell,t}^{m,j}(\tau)}R_{t+\ell},
\end{aligned}
\end{equation*}
where $D_{0,t}^{m,j}=E_{1,t}^{m,j}$ and
$D_{1,t}^{m,j}=E_{2,t}^{m,j}$. The pathwise discrepancy from the standard
action-level return is
\begin{equation*}
\begin{aligned}
\Delta_{\gamma_w}^{m,j}(\tau)
&\triangleq G_t(\tau)
-\widetilde G_{\gamma_w}(\tau\mid\bar z_t^{m,j})\\
&=\sum_{\ell=0}^{\infty}\gamma_a^{\ell}
\bigl(1-\gamma_w^{D_{\ell,t}^{m,j}(\tau)}\bigr)R_{t+\ell}\\
&=(1-\gamma_w^{E_1})R_t
+\gamma_a(1-\gamma_w^{E_2})R_{t+1}+\cdots.
\end{aligned}
\end{equation*}
Making the dependence on the intra-action discount
explicit, taking conditional expectations of the pathwise identity gives
\begin{equation*}
\begin{aligned}
\widetilde Q_{\gamma_w}^{\bm\pi}(z_t^{m,j},w_t^{m,j})
&=\widetilde Q_{1}^{\bm\pi}(z_t^{m,j},w_t^{m,j})
-\mathbb E_{\tau\sim p_{\bm\pi}
(\cdot\mid z_t^{m,j},w_t^{m,j})}
\bigl[\Delta_{\gamma_w}^{m,j}(\tau)\bigr],\\
\widetilde V_{\gamma_w}^{\bm\pi}(\bar z_t^{m,j})
&=\widetilde V_{1}^{\bm\pi}(\bar z_t^{m,j})
-\mathbb E_{\tau\sim p_{\bm\pi}
(\cdot\mid\bar z_t^{m,j})}
\bigl[\Delta_{\gamma_w}^{m,j}(\tau)\bigr].
\end{aligned}
\end{equation*}
These equations compare $\gamma_w$ with $\gamma_w=1$
at the same prefix under the same fixed policy. At aligned complete-action
boundaries,
\begin{equation*}
\widetilde Q_{1}^{\bm\pi}(s,\bm a)=Q^{\bm\pi}(s,\bm a),
\qquad
\widetilde V_{1}^{\bm\pi}(s)=V^{\bm\pi}(s).
\end{equation*}
Standard action-level values are recovered only at
aligned boundaries; arbitrary partial-token prefixes remain conditional. When
$\gamma_w<1$, length weighting may change the optimal policy
\citep{wen2024reinforcinglanguageagentspolicy}. For~nonnegative rewards,
$\Delta_{\gamma_w}^{m,j}$ is nondecreasing as $\gamma_w$ decreases. At
$\gamma_w=1$, it vanishes pathwise; averaging over policy randomness and the
transition kernel $P$ preserves boundary values and the optimal objective.
Appendix~\ref{appendix:token-level-suboptimality} proves these claims under
the stated integrability and common-policy-class assumptions. We set
$\gamma_w=1$ while retaining token-level TD errors, advantages, and PPO
updates.
\endgroup

\subsection{Implementations}
\label{sec:marft-methods:implementations}

\subsubsection{Base Algorithm}
\label{sec:marft-methods:implementation:action-level}

\reviewchange{The detailed algorithm implementation of MARFT is illustrated in Algorithm~\ref{alg:marft}.} For the design of the value function $V_\phi$, we base it on a frozen transformer module, which processes the sequence information and outputs the final hidden state vector of the sequence, and add an extra trainable multilayer perceptron (MLP) with parameters $\phi$ to map the hidden vector to a specific value. A common instantiation of the transformer module can be the base model of an agent. For the design of the agents, we usually resort to LoRA fine-tuning to avoid overwhelming pressure on computing resources. If the agents share the same base model, we can assign separate but same-structured LoRA adapters to them, but if the agents have different base models, even different modalities, we then have to give them distinguished LoRA adapters.

\begin{algorithm}[htbp]
    \caption{Multi-Agent Reinforcement Fine-Tuning (MARFT)}
    \label{alg:marft}
    \SetAlgoLined
    \KwIn{Agent population $n$, agent profiles $\text{inst}^i$, the initial joint policy $\bm\pi_{\bm\theta_0}$ with parameters $\theta_0^i$ for each policy $\pi^{i}_{\theta^i}$, initial parameters $\phi_0$ for the critic network $V_\phi$, hyper-parameters including maximum interaction steps $T$ in one roll-out, clip parameter $\epsilon$, discount factor $\gamma$, GAE $\lambda$, etc.}
    \KwOut{Optimized joint policy $\bm\pi$ and critic network $\phi$}
    initialize policy $\bm\pi_{\bm\theta} \gets \bm\pi_{\bm\theta_0}$, critic network $V_{\phi} \gets V_{\phi_0}$ and buffer $\mathcal{D} \gets \emptyset$\\
    \For{$\text{episode}=0,\dots$}{
        \For{$t=0$, \dots, $T-1$}{
            collect $s_t$ \\
            \For{$i=1$, \dots, $n$}{
                generate action $a^i \sim p_{\pi^{i}_{\theta^i}}(a^i|s,\bm{a}^{1:i-1})$
            }
            $s_{t+1} \sim \mathcal{T}(\cdot|s_t,\bm{a}_t)$\\
            $r_t = R(s_t,\bm{a}_t)$\\
            $\mathcal{D} \gets \mathcal{D} \cup \{\left( s_t, \bm{a}_t, r_t, s_{t+1}\right)\}$\\
        }
        compute advantage estimate $\hat{A}$ via GAE and compute value network target $\hat{V}$ in $\mathcal{D}$\\
        \For{$\text{n epoches}$}{
            sample a batch $\mathcal{B} = \{( s_t, \bm{a}_t, r_t, s_{t+1},\hat{A},\hat{V})\}\sim\mathcal{D}$\\
            update $\phi$ by minimizing $\sum_{n=1}^N \left\| V_\phi(s_n)- \hat{V}_n \right\|^2$\\
            update $\bm{\theta}$ by maximizing the objective \eqref{eq:marft-objective}\\
        }
    }
\end{algorithm}
For the implementation of action normalization, we can directly count the token length or word length as $Z$ and divide the sum of the log probabilities of generated tokens by $Z$. 
\begin{align}
    \label{eq:twosome-normalization}
    {\text{Norm}}\left(\text{log}P_{\pi^{i}_{\theta^i}}\left(a^{i}|s,\bm{a}^{1:i-1} \right)\right) = \frac{\text{log}P_{\pi^{i}_{\theta^i}}\left(a^{i}|s,\bm{a}^{i-1}\right)}{Z}.
\end{align}
To implement agent-by-agent update, this sequential update mechanism involves updating only one policy per training epoch and cycling through the policies alternatively.

\subsubsection{Token-level Adaptations}
\label{sec:marft-methods:implementation:token-level}

\begingroup
\tokenrevision
For token-level MARFT, the outer loop of Algorithm~\ref{alg:marft} stores
token logits and conditional token values during rollout. We set
$\gamma_w=1$, apply $\gamma_a$ only at environment boundaries, and use the
token-level backups above to construct critic targets and TD errors.
GAE~\citep{schulman2018highdimensionalcontinuouscontrolusing} estimates the
advantages used by the clipped PPO objective at token granularity. The
return-consistency result preserves the action-level objective under the
stated assumptions; it does not imply exact convergence of finite-sample PPO.
\endgroup

\begingroup
\section{Experiments}
\label{sec:experiments}

We evaluate whether MARFT improves fixed two- to four-agent math and coding workflows relative to IPPO, and examine policy sharing, agent count, and role prompting. This IPPO comparator follows the independent-learning structure used in related post-training work, including MAPoRL \citep{park2025maporlmultiagentpostcotrainingcollaborative}. Our implementation is built on the open-source AReaL codebase \citep{fu2025areal}.

\begin{table}[!t]
\centering
\caption{Performance comparison of MARFT and IPPO. Under MARFT, PS indicates that actors share policy parameters, whereas w/o PS assigns a separate policy adapter to each actor. Math rows report accuracy in percentage points, while coding rows retain each benchmark's reported score. Values are means across five stochastic decoding seeds for one trained checkpoint in each configuration. Dispersion is omitted from the displayed table. Bold marks the highest available mean within each LaMAS--benchmark setting.}
\label{tab:evaluation_results}
\begingroup
\small
\setlength{\tabcolsep}{1pt}
\renewcommand{\arraystretch}{1.08}
\begin{tabular*}{\linewidth}{@{\extracolsep{\fill}}l|cccc|cccc|cccc@{}}
\hline
\multirow{3}{*}{\textbf{Benchmark}} &
\multicolumn{4}{c|}{\textbf{Duo}} &
\multicolumn{4}{c|}{\textbf{Trio}} &
\multicolumn{4}{c}{\textbf{Quad}} \\
\cline{2-5}\cline{6-9}\cline{10-13}
& \multirow{2}{*}{\textbf{Van.}} & \multirow{2}{*}{\textbf{IPPO}} & \multicolumn{2}{c|}{\textbf{MARFT}}
& \multirow{2}{*}{\textbf{Van.}} & \multirow{2}{*}{\textbf{IPPO}} & \multicolumn{2}{c|}{\textbf{MARFT}}
& \multirow{2}{*}{\textbf{Van.}} & \multirow{2}{*}{\textbf{IPPO}} & \multicolumn{2}{c}{\textbf{MARFT}} \\
\cline{4-5}\cline{8-9}\cline{12-13}
& & & \textbf{PS} & \textbf{w/o PS}
& & & \textbf{PS} & \textbf{w/o PS}
& & & \textbf{PS} & \textbf{w/o PS} \\
\hline
DeepCoder     & 45.28  & 46.95        & \textbf{52.68} & 50.02 & 50.38  & 51.38 & \textbf{55.18} & 53.97 & 47.16  & 51.30        & \textbf{57.02} & 56.33 \\
LiveCodeBench & 169.19 & 176.93 & 180.23 & \textbf{183.98} & 160.51 & 162.72      & 167.18 & \textbf{170.22} & 152.09 & 156.48       & 164.92 & \textbf{167.52} \\
CodeForces    & 24.04  & 24.28        & 26.59 & \textbf{29.47} & 24.00  & 24.06        & 27.68 & \textbf{27.97} & 20.50  & 21.88        & \textbf{28.05} & 25.07 \\
AIME2024      & 2.00   & 3.33  & \textbf{10.00} & 6.67 & 2.00   & 4.00        & \textbf{6.67} & \textbf{6.67} & 2.00   & 2.67        & \textbf{6.67} & 3.33 \\
MATH500       & 43.20  & 49.48 & \textbf{52.00} & 51.20 & 45.76  & 49.12       & 50.00 & \textbf{51.00} & 46.64  & 49.76       & 50.60 & \textbf{51.60} \\
Minerva Math  & 13.46  & 16.32 & 18.01 & \textbf{18.38} & 15.81  & 16.76       & \textbf{17.28} & 16.91 & 15.29  & 15.81       & \textbf{17.65} & 16.54 \\
DeepScaleR & 20.15 & 22.46    & \textbf{23.31} & 22.79 & 20.79  & 22.54 & \textbf{23.31} & 23.16 & 21.23  & 22.24       & \textbf{22.89} & 22.72 \\
OlympiadBench & 16.08  & 18.31       & \textbf{21.22} & 20.33 & 17.60  & 19.05 & 20.18 & \textbf{20.33} & 16.60  & 18.37       & 19.58 & \textbf{19.73} \\
\hline
\textbf{Average} & 41.68 & 44.76 & \textbf{48.00} & 47.86 & 42.11 & 43.70 & 45.94 & \textbf{46.27} & 40.19 & 42.31 & \textbf{45.92} & 45.36 \\
\hline
\end{tabular*}
\endgroup
\end{table}

\subsection{Experiment Setups}
\label{sec:experiments:setups}

\noindent\textbf{Benchmarks and Rewards.}
We train on DeepCoder Preview for coding and DeepScaleR for mathematics \citep{deepcoder2025,tan2026deepscaler}. We evaluate coding on LiveCodeBench and CodeForces \citep{jain2025livecodebench,penedo2025codeforces}, and mathematics on AIME2024, MATH500, Minerva Math, DeepScaleR, and OlympiadBench \citep{AIME2024,hendrycks2021measuringmathematicalproblemsolving,lightman2024letsverify,lewkowycz2022minerva,he-etal-2024-olympiadbench}. For DeepCoder and DeepScaleR, the evaluation examples are separately held out and do not overlap with the corresponding training sets. Training rewards are binary correctness for mathematics and a test-case pass rate or score for coding.

\noindent\textbf{Base Model and LaMAS Configurations.}
We initialize the mathematics experiments from Qwen2.5-1.5B-Instruct and the coding experiments from Qwen2.5-3B-Instruct \citep{qwen2}. We evaluate three LaMAS configurations: \Duo{} uses Planner~$\rightarrow$~Solver; \Trio{} adds a Verifier; and \Quad{} further inserts a Reflector before the Verifier. Each role receives a concise system prompt, and the final agent produces the answer.

\noindent\textbf{LoRA Fine-Tuning Configurations.}

We use LoRA with rank 32, alpha 16, and all-linear target modules, so each adapter contains about 2\% as many tunable parameters as the base model. We evaluate MARFT with and without policy parameter sharing. With sharing, all agents use one policy LoRA adapter; without it, each agent has a separate adapter. Both configurations use one centralized-critic adapter. Thus, for a LaMAS with $n$ agents, the two configurations train two and $n+1$ adapters, respectively, whereas IPPO trains $2n$ adapters for $n$ independent policies and critics. Appendix~\ref{app:experiment-hyperparameters} reports the decisive optimization and generation settings for every released experiment configuration.

\subsection{Comprehensive Comparison Results}
\label{sec:experiments:comparison}

\subsubsection{MARFT achieves higher mean evaluation scores} Table~\ref{tab:evaluation_results} compares vanilla LaMAS, IPPO, and MARFT with and without parameter sharing across the complete suite. Across all 24 settings, both MARFT configurations exceed IPPO. Parameter sharing yields the higher score in thirteen settings, while training without parameter sharing is higher in ten, with one tie. The unweighted descriptive means for IPPO, MARFT with parameter sharing, and MARFT without parameter sharing are respectively 44.76, 48.00, and 47.86 for \Duo{}; 43.70, 45.94, and 46.27 for \Trio{}; and 42.31, 45.92, and 45.36 for \Quad{}. On \Quad{} coding benchmarks, MARFT with parameter sharing exceeds IPPO by 5.72 on DeepCoder, 8.44 on LiveCodeBench, and 6.17 on CodeForces. These comparisons are descriptive because dispersion is unavailable, but they show that the relative effect of parameter sharing varies across the reported benchmarks rather than favoring one configuration uniformly.

\FloatBarrier

\subsubsection{MARFT exhibits more stable training improvements} Figure~\ref{fig:training_reward_figures} compares the training dynamics of MARFT and IPPO on DeepScaleR and DeepCoder across \Duo{}, \Trio{}, and \Quad{}. MARFT, particularly without parameter sharing, maintains a more sustained upward reward trend across both domains. Local oscillations are expected under stochastic on-policy optimization, but the MARFT curves generally avoid persistent reward degradation and reach competitive or higher terminal rewards. On DeepCoder \Duo{}, IPPO drops sharply near step 550 before recovering; both MARFT variants remain stable. This supports the stability motivation of predecessor-conditioned PPO, not a strict monotonicity guarantee for finite-sample PPO.

\begin{figure}[!htbp]
    \centering
    \begin{subfigure}{0.98\linewidth}
        \centering
        \includegraphics[width=\linewidth]{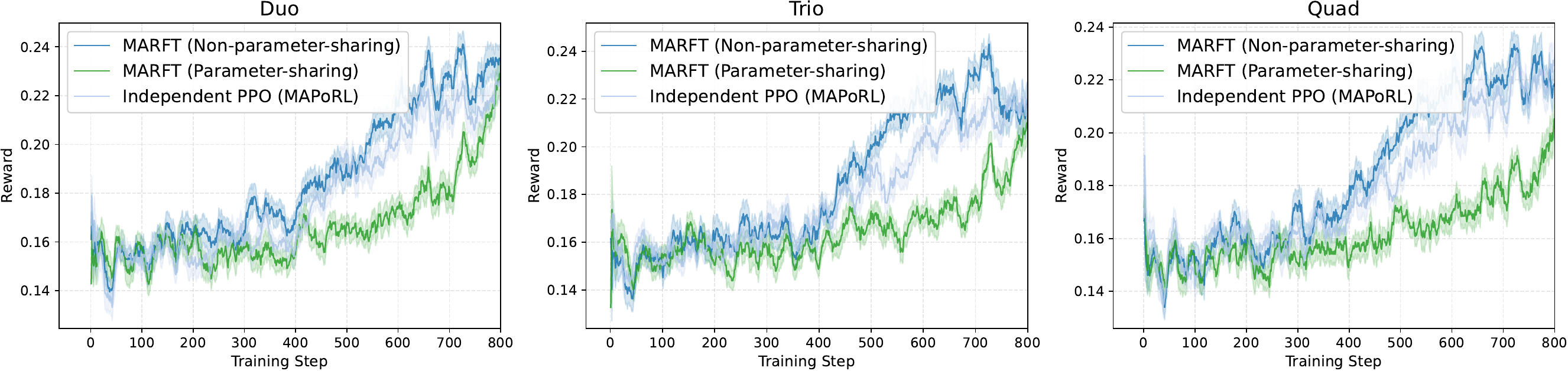}
        \caption{Training reward on DeepScaleR.}
        \label{fig:ablation_algo}
    \end{subfigure}

    \begin{subfigure}{0.98\linewidth}
        \centering
        \includegraphics[width=\linewidth]{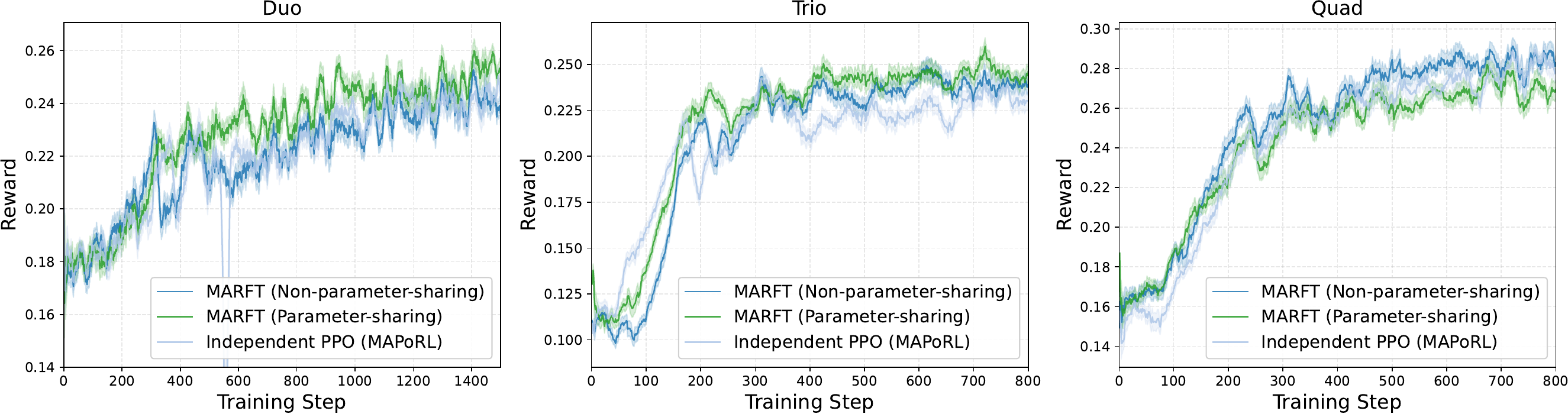}
        \caption{Training reward on DeepCoder.}
        \label{fig:ablation_algo_deepcoder}
    \end{subfigure}
    \caption{Training rewards when fine-tuning LaMAS using MARFT and IPPO across different environments and agent scales. In the legend, \textit{Independent PPO (MAPoRL)} denotes the IPPO comparator used throughout this paper; the MAPoRL attribution indicates that its optimization backbone follows MAPoRL's independent PPO-style structure.}
    
    \label{fig:training_reward_figures}
\end{figure}
\FloatBarrier

An interesting phenomenon in both Table~\ref{tab:evaluation_results} and Figure~\ref{fig:training_reward_figures} is that increasing the number of agents does not yield a consistent performance gain: some benchmark scores increase, whereas others decrease. For MARFT with parameter sharing, the mixed-scale descriptive means are 48.00 for \Duo{}, 45.94 for \Trio{}, and 45.92 for \Quad{}, which provides the same directional observation but is not a formal scaling statistic. A plausible explanation is diminishing returns in a shared-backbone, role-asymmetric multi-agent system. Recent information-theoretic analysis suggests that performance depends on the number of effective channels rather than raw agent count. Here, effective channels refer to independent, non-redundant reasoning paths produced by the agents \citep{yang2026understandingagentscalingllmbased}. Because our \Duo{}, \Trio{}, and \Quad{} configurations use the same base model, additional agents may generate highly correlated trajectories that inject redundant evidence instead of complementary information. This explanation remains a hypothesis because the present experiments do not directly measure reasoning-path diversity.

Crucially, although naive agent scaling provides limited gains in this shared-backbone setting, both MARFT configurations, one with and one without parameter sharing, have higher reported evaluation means than IPPO at every tested scale and benchmark. Figure~\ref{fig:training_reward_figures} additionally compares the two configurations during training. The available evidence supports the complete MARFT system, while their mixed relative performance prevents a general conclusion about whether sharing policy parameters is preferable.

\FloatBarrier
\subsection{Ablation on Role Specification}
\label{sec:experiments:self-evolution}
We finally examine how removing explicit role descriptions affects MARFT training on DeepScaleR. In this anonymous setting, agents retain generic position-indexed system prompts but receive no role-specific instructions, allowing MARFT to optimize their interaction without prescribing a division of labor in advance. Figure~\ref{fig:self-evolution-ablation} compares these anonymous variants with their preconfigured counterparts across \Duo{}, \Trio{}, and \Quad{}, both with and without parameter sharing.

Across all six comparisons, the anonymous variants achieve higher rewards over most of training, with the clearest margins in the early and middle stages. The gaps narrow late in several settings, especially without parameter sharing, but the overall pattern remains favorable to anonymous initialization. A plausible explanation is that general-purpose base models incur an adaptation cost when first required to follow narrowly specified roles. Removing those specifications may allow MARFT to optimize the observed training objective more directly, without first fitting a prescribed behavioral partition. Thus, under the tested initialization, explicit role prompting can hinder early optimization.

\begin{figure}[!htbp]
    \centering
    \includegraphics[width=0.98\linewidth]{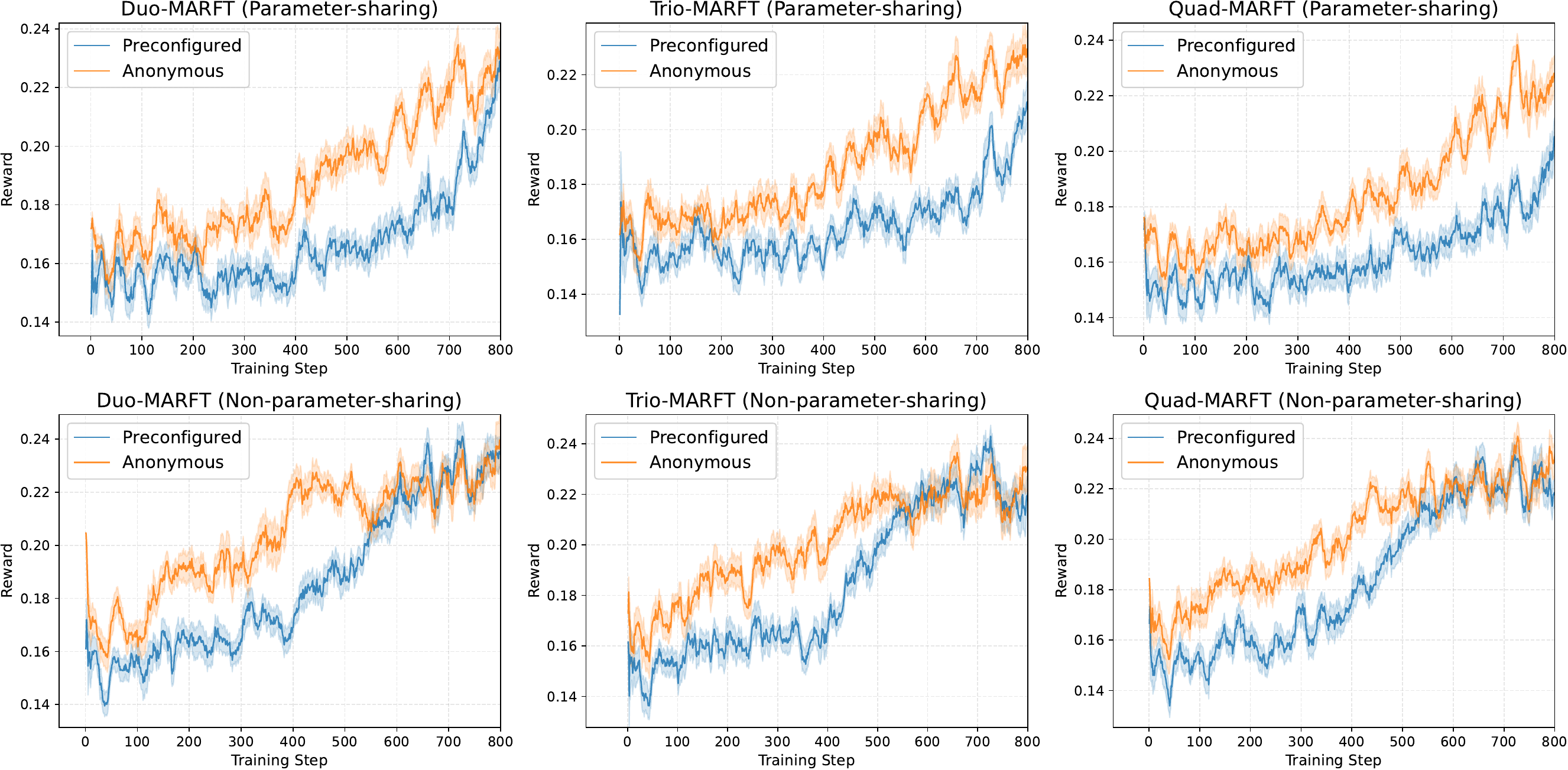}
    \caption{Training rewards on DeepScaleR~under preconfigured and anonymous role prompting.}
    
    \label{fig:self-evolution-ablation}
\end{figure}
\FloatBarrier
\endgroup

\section{Perspectives}
\label{sec:perspectives}

The preceding experiments evaluate fixed, sequential math and coding workflows with two to four agents. From this section onward, we discuss broader application perspectives and challenges that extend beyond those evaluated settings.

\paragraph{Powerful Capabilities for Solving Complex Agentic Tasks}
Agentic tasks are significantly more complex than traditional language tasks, requiring capabilities such as autonomous decision-making, collaboration, and adaptive reasoning. While a standalone LLM-based agent may lack the breadth of skills needed to address these multifaceted challenges, MARFT empowers LaMAS to overcome these limitations. By leveraging RFT, MARFT enables the efficient decomposition of complex instructions into executable sub-tasks, distributing them among specialized agents within the LaMAS framework. This allows heterogeneous LLM-based agents to contribute their unique intelligence while dynamically sharing goals, negotiating strategies, and aligning team objectives through natural language interactions. MARFT further enhances this process by optimizing both individual agent policies and collective system performance, ensuring efficient coordination and adaptability. For example, in a logistics scenario, MARFT fine-tunes agents to collaboratively solve the task of "delivering emergency supplies" by dynamically assigning sub-tasks such as route planning, inventory allocation, and risk assessment to specialized agents, thereby maximizing the system's overall effectiveness.

\paragraph{Scalability and Robust Generalization} MARFT offers significant scalability. Agentic multi-agent problems are highly flexible, allowing for various ways to decompose tasks and design corresponding multi-agent systems. As the agent population, environmental complexity, and task scale increase, the key challenge is to maintain efficient and stable learning performance and coordination capabilities. MARFT not only enhances the overall performance of the multi-agent system but also increases the intelligence of each agent by enabling more dynamic role-playing. Even if the system organization changes significantly, the problem can be re-decomposed to allow agents to function within a multi-agent system similar to the one they were trained in.

\paragraph{Enhanced Privacy and Federated Learning} In a unified multi-agent system, agents do not have access to each other's local data. Instead, they contribute their unique intelligence to the system through their actions, thereby enhancing the system's collective intelligence. This setup aligns with many real-world scenarios where entities are willing to share their powerful agents but are highly concerned about data privacy. While this resembles federated learning, MARFT places greater emphasis on improving system performance through better collaboration among agents.

\paragraph{Integration with Blockchain Technology} MARFT's decentralized nature and privacy-preserving capabilities make it a natural fit for blockchain. In blockchain systems, MARFT can enable secure and efficient multi-agent collaboration without requiring agents to share sensitive data. For example, in smart contract execution, MARFT can coordinate multiple agents to validate transactions, manage decentralized autonomous organizations (DAOs), or optimize resource allocation in decentralized finance (DeFi) platforms. The ability of MARFT to dynamically adapt to changing environments and agent populations ensures robust performance in the unpredictable and adversarial settings typical of blockchain ecosystems. Additionally, MARFT's emphasis on preserving pretrained capabilities while adapting to new tasks aligns well with the need for agents to maintain trustworthiness and reliability in blockchain-based systems. This integration opens new avenues for enhancing transparency, security, and efficiency in blockchain applications while leveraging the collective intelligence of LaMAS.
\section{Open Problems}
\label{sec:open-problems}

Despite the advantages and potential of MARFT paradigm, both academia and industry face numerous challenges that impede the development of more effective MARFT algorithms.

\paragraph{Lack of Dynamic Environments for Training} A significant challenge in advancing MARFT is the absence of well-established, easy-to-implement, and scalable dynamic interactive environments designed for solving agentic tasks. There are two primary difficulties in this regard. First, creating such environments with complex agentic tasks demands substantial engineering expertise, comparable to developing fundamental environments like VirtualHome \citep{puig2018virtualhome} or Overcooked \citep{NEURIPS2019_f5b1b89d} from scratch. Second, designing reward feedback mechanisms for multi-agent systems in highly dynamic environments with complex agentic tasks is exceedingly challenging. The reward signals can be multi-dimensional, and balancing the weights among different goals is a delicate task. Although there are several "dynamic" benchmarks available, such as the Berkeley Function Calling Leaderboard \citep{berkeley-function-calling-leaderboard,patil2023gorilla}, ToolSandBox \citep{lu2024toolsandboxstatefulconversationalinteractive}, and GAIA \citep{mialon2024gaia}, converting these into dynamic environments that support MARL training remains an unresolved issue.

\paragraph{Low Sample Efficiency and Lack of High-Quality Synthetic Data} RL, particularly on-policy RL, is known for its low sample efficiency. On-policy algorithms such as PPO and TRPO require frequent switching between sampling trajectories and training. Their trust-region motivation should be distinguished from the behavior of practical finite-sample clipped PPO, which has no unconditional monotonic-improvement guarantee. This process is especially problematic and time-consuming when applied to LLMs. Potential solutions may include adopting ideas like Dyna \citep{sutton1991dyna} to enhance sample efficiency or developing algorithms that effectively utilize stale trajectories. Another issue stemming from low sample efficiency is the scarcity of high-quality synthetic data. Effective multi-agent trajectories need to achieve high success rates while also demonstrating efficient communication and coordination. Currently, the field lacks such data for multi-agent cold starts.

\paragraph{Lack of an Established MARFT Framework} Developing engineering solutions that integrate LLMs and MARL is highly challenging. While there exist user-friendly, efficient, and scalable frameworks such as verl~\citep{sheng2024hybridflow}, OpenRLHF~\citep{hu2024openrlhf} for LLM fine-tuning and MALib~\citep{JMLR:v24:22-0169}, ReMA~\citep{wan2025remalearningmetathinkllms}, PettingLLMs~\citep{zhao2025stronger} for MARL training, a comprehensive framework specifically designed for MARFT is notably absent. The creation of such a framework would require significantly more effort than developing either of the aforementioned frameworks individually. This is because it must effectively integrate both LLMs and MARL components, each of which presents its own set of complexities. While this article introduces a MARFT framework, it is still in its early stages and offers significant room for further development and refinement.

\paragraph{Lack of a Unified Communication Mechanism or Protocol} Recent advancements have led to the development of numerous agent communication protocols~\citep{yang2025surveyaiagentprotocols}, such as MCP (Model Context Protocol)~\citep{model_context_protocol}, A2A (Agent-to-Agent)~\citep{a2a_protocol}, and ANP (Agent Network Protocol) ~\citep{anp_protocol}. These protocols are essential for enabling effective collaboration among agents and facilitating interactions with users. They also address the critical need for agents to contribute to shared goals while preserving their own data and knowledge, reflecting the future vision of a decentralized, privacy-preserving agent ecosystem. However, the proliferation of these protocols has resulted in a fragmented landscape, lacking a unified and efficient communication pipeline. This fragmentation may hinder the seamless integration and interoperability of diverse agents, particularly in dynamic and heterogeneous environments typical of LaMAS. The absence of a standardized communication framework not only complicates system design and scalability but also introduces inefficiencies and potential bottlenecks in real-world applications.

\section{Conclusion}
\label{sec:conclusion}

In this article, we propose a novel paradigm termed \textbf{Multi-Agent Reinforcement Fine-Tuning (MARFT)}, together with a brand-new MG called Flex-MG to better align with the LaMAS optimization in realistic scenarios, and present a universal algorithmic framework, MARFT, that integrates RFT, LaMAS, and MARL to advance the capabilities of LaMAS systems. MARFT addresses several fundamental challenges in adapting conventional MARL to LaMAS, including asynchronous agent interactions, profile-aware architecture design, heterogeneity across agents, and the dynamic organization of multi-agent systems. Through the fusion of RFT with sequential decision-making paradigms, trust-region optimization techniques, and token-level adaptation mechanisms, MARFT enables efficient coordination and policy refinement while preserving the foundational capabilities of pretrained LLMs.

\reviewchange{Empirically, across the evaluated \Duo{}, \Trio{}, and \Quad{} math and coding settings, both MARFT configurations report higher mean evaluation scores than IPPO in all 24 benchmark--team-size settings, while the variant without parameter sharing shows the most consistent advantage in final training reward. The DeepScaleR role-specification ablation further indicates that removing prescribed role prompts is associated with higher observed training rewards, particularly in the early and middle stages.} These findings suggest that MARFT holds significant promise for agentic tasks requiring complex, well-organized, distributed reasoning and acting.

The paradigm's intrinsic scalability, privacy-preserving characteristics, and alignment with decentralized system architectures make it a compelling foundation for real-world deployment in domains ranging from logistics and collaborative robotics to blockchain-based ecosystems. Nevertheless, numerous theoretical and practical challenges remain. Key open problems include the development of dynamic environment designs with agentic tasks tailored for LaMAS, improvement of sample efficiency, and the absence of standardized and well-established frameworks for MARFT and a unified multi-agent communication protocol.

Future research directions may include the creation of unified benchmarks and development toolkits for MARFT, the construction of high-quality synthetic datasets to simulate multi-agent LLM interactions, and the design of hybrid learning strategies to boost data efficiency, etc. By addressing these open challenges, we envision MARFT as a stepping stone toward realizing generalizable, human-aligned, and collaboratively adaptive agent systems, paving the way for scalable AGI capable of tackling complex open-world problems with robust agentic behavior and human-like adaptability.


\bibliography{main}
\bibliographystyle{rlc}

\newpage
\appendix

\begingroup
\tokenrevision
\section{Token-Level Return Consistency}
\label{appendix:token-level-suboptimality}
\label{app:token-level-proof}

This appendix states the assumptions behind the token-level Bellman
construction in Section~\ref{sec:marft-methods:methodology} and proves the consistency result used
by MARFT. Here $\gamma_a=\gamma$ is the Flex-MG discount and
$0\leq\gamma_a,\gamma_w\leq1$. The argument is pathwise: it first compares the return assigned to a
fixed tokenized trajectory with its action-level return, and then takes policy
expectations. This avoids identifying a value conditioned on a partial token
prefix with an unconditional state value.

\subsection{Discount Schedule and Assumptions}

Each joint action $\bm a_t^{1:n}$ is a finite collection of token sequences,
with $|a_t^k|\geq 1$. Intermediate token rewards are zero. A factor
$\gamma_w$ is applied after an interior token transition, no $\gamma_w$
factor is applied when execution crosses an agent boundary, and the environment
discount $\gamma_a$ is applied once between successive environment rewards.
The environment transition remains the Flex-MG kernel
$P(\cdot\mid s_t,\bm a_t)$; all expectations below therefore include both
policy and transition randomness. The same EOS/truncation convention and action boundaries are used in
the token-level and action-level descriptions. Powers with exponent zero
represent the empty product, including $\gamma_w^0=1$ when $\gamma_w=0$.
For every admissible policy and state/action or token-history conditional law
used below, we assume
$\mathbb E[\sum_{\ell\geq0}\gamma_a^\ell|R_{t+\ell}|]<\infty$, which is
satisfied, for example, by a uniformly bounded finite horizon with finite
conditional first moments of rewards, or by uniformly bounded rewards with
$0\leq\gamma_a<1$. Thus, the returns below are absolutely
integrable.

Starting immediately after agent $m$ emits token $j$, the number of
within-action discount factors before the current reward is
\begin{equation*}
\begin{aligned}
E_{1,t}^{m,j}(\bm a_t)
&=(|a_t^m|-j)
  +\sum_{k=m+1}^{n}(|a_t^k|-1)\\
&=\sum_{k=m}^{n}|a_t^k|-j-(n-m).
\end{aligned}
\end{equation*}
The first term counts the remaining interior-token transitions of agent $m$.
For each later agent, $|a_t^k|-1$ transitions receive $\gamma_w$ because the
cross-agent boundary itself is undiscounted.

Define the number of intra-action transitions in a complete joint action by
\begin{equation*}
C(\bm a)\triangleq\sum_{k=1}^{n}(|a^k|-1).
\end{equation*}
For a fixed next joint action $\bm a_{t+1}$, the exponent before the next
environment reward is therefore
\begin{equation*}
\begin{aligned}
E_{2,t}^{m,j}(\bm a_t,\bm a_{t+1})
&=E_{1,t}^{m,j}(\bm a_t)+C(\bm a_{t+1})\\
&=\sum_{k=m}^{n}|a_t^k|
 +\sum_{k=1}^{n}|a_{t+1}^k|+m-j-2n.
\end{aligned}
\end{equation*}
In particular, $E_2$ depends on the candidate next action through its generated
length. It cannot generally be factored outside a policy expectation or
maximization.

\subsection{Pathwise Token-Level Return}

Fix a complete trajectory
$\tau=(s_t,\bm a_t,R_t,s_{t+1},\bm a_{t+1},R_{t+1},\ldots)$ with
$\sum_{\ell\geq0}\gamma_a^\ell|R_{t+\ell}|<\infty$, as holds almost
surely under the integrability assumption, and a post-token history
$\bar z_t^{m,j}$. For the sampled token at any pre-token
history $z_t^{m,j}$, we use the convention
\begin{equation*}
\widetilde G_{\gamma_w}(\tau\mid z_t^{m,j})
=\widetilde G_{\gamma_w}(\tau\mid\bar z_t^{m,j}).
\end{equation*}
Token emission carries no additional factor; interior discounting is applied
between successive post-token histories. Consequently, for a fixed policy,
\begin{equation*}
\begin{aligned}
\widetilde V_{\gamma_w}^{\bm\pi}(z_t^{m,j})
&=\mathbb E_{w\sim\pi^m(\cdot\mid z_t^{m,j})}
 [\widetilde Q_{\gamma_w}^{\bm\pi}(z_t^{m,j},w)],\\
\widetilde V_{\gamma_w}^{\bm\pi}(\bar z_t^{m,j})
&=\widetilde Q_{\gamma_w}^{\bm\pi}(z_t^{m,j},w_t^{m,j}).
\end{aligned}
\end{equation*}
For $\ell\geq0$, let
\begin{equation*}
D_{\ell,t}^{m,j}(\tau)
\triangleq
E_{1,t}^{m,j}(\bm a_t)
+\sum_{r=1}^{\ell}C(\bm a_{t+r}),
\end{equation*}
where the empty sum is zero. Thus
$D_{0,t}^{m,j}=E_{1,t}^{m,j}$ and
$D_{1,t}^{m,j}=E_{2,t}^{m,j}$. Unrolling the two-scale discount
schedule along the fixed sampled tokens and states gives the following
pathwise return:
\begin{equation}
\widetilde G_{\gamma_w}(\tau\mid\bar z_t^{m,j})
=
\sum_{\ell=0}^{\infty}
\gamma_a^{\ell}
\gamma_w^{D_{\ell,t}^{m,j}(\tau)}
R_{t+\ell}.
\label{app:eq:token-return}
\end{equation}
Equation~\eqref{app:eq:token-return} is a return identity along a fixed
trajectory, rather than an equality between a partial-prefix value and the
value of a preselected complete action. When $\gamma_w<1$, different output
lengths induce different weights on otherwise corresponding environment
rewards and can therefore change the optimized objective.

Setting $\gamma_w=1$ makes every intra-action factor equal to one. Hence, for
every complete trajectory with absolutely convergent return, and therefore
almost surely under the stated assumptions,
\begin{equation}
\widetilde G_{1}(\tau\mid\bar z_t^{m,j})
=
\sum_{\ell=0}^{\infty}\gamma_a^{\ell}R_{t+\ell}
=G_t(\tau).
\label{app:eq:pathwise-consistency}
\end{equation}
This equality holds before taking any expectation and is unaffected by whether
$P(\cdot\mid s,\bm a)$ is deterministic or stochastic.

\subsection{Consequences for Policy and Optimal Values}

Taking expectations of Eq.~\eqref{app:eq:pathwise-consistency} under any fixed
joint policy $\bm\pi$ and the Flex-MG transition kernel shows that the
tokenized and action-level returns agree at aligned action boundaries. In
particular, let $z_t^{1,1}=(s_t,\emptyset,\emptyset)$ denote the history before
the first token of the joint action; choosing that first token incurs no
preceding $\gamma_w$ factor. Define the complete-action token value by
\begin{equation*}
\widetilde Q_{\gamma_w}^{\bm\pi}(s,\bm a)
\triangleq
\mathbb E\!\left[
\widetilde G_{\gamma_w}(\tau\mid z_t^{1,1})
\,\middle|\,s_t=s,\bm a_t=\bm a
\right].
\end{equation*}
and
\begin{equation*}
\widetilde V_{\gamma_w}^{\bm\pi}(s)
\triangleq
\mathbb E_{\bm a\sim\bm\pi(\cdot\mid s)}
\left[\widetilde Q_{\gamma_w}^{\bm\pi}(s,\bm a)\right].
\end{equation*}
Here the current complete action is fixed, and future actions follow
$\bm\pi$, as in the usual action-value definition; a current action need
not be sampled with positive probability under that policy. The boundary
value starts before the first token and therefore includes discounting
within the fixed current action when $\gamma_w<1$; it is distinct from
the value starting after its final token. Then, for every admissible
complete joint action $\bm a$,
\begin{equation}
\widetilde Q_{\gamma_w=1}^{\bm\pi}(s,\bm a)
=Q^{\bm\pi}(s,\bm a),
\qquad
\widetilde V_{\gamma_w=1}^{\bm\pi}(s)
=V^{\bm\pi}(s).
\label{app:eq:policy-consistency}
\end{equation}
The action-level Bellman equation retains its transition expectation,
\begin{equation*}
Q^{\bm\pi}(s,\bm a)
=
R(s,\bm a)
+\gamma_a
\mathbb E_{s'\sim P(\cdot\mid s,\bm a)}
\left[V^{\bm\pi}(s')\right].
\end{equation*}
Because Eq.~\eqref{app:eq:policy-consistency} holds for every policy, optimizing
over the same complete-action policy class preserves the optimal objective at
action boundaries. A value after an arbitrary partial prefix remains
conditioned on that prefix and need not equal the unconditional
$V^{\bm\pi^\star}(s)$.

Thus, under the zero-intermediate-reward, boundary-discount, tokenization, and
transition-expectation assumptions above,
\begin{equation*}
\boxed{\gamma_w=1}
\end{equation*}
is a general sufficient condition for removing the artificial
token-length discount while retaining token-granularity TD estimation and PPO
optimization. It is not a necessary condition on a particular trajectory when
$\sum_{\ell\geq0}\gamma_a^\ell
(1-\gamma_w^{D_{\ell,t}^{m,j}})R_{t+\ell}=0$; sufficient degenerate examples
include all reward terms being zero or every nonzero reward term having
$D_{\ell,t}^{m,j}=0$. For $0\leq\gamma_w\leq1$ and a fixed trajectory with
nonnegative rewards, the difference
$G_t-\widetilde G_{\gamma_w}
=\sum_{\ell\geq0}\gamma_a^\ell
(1-\gamma_w^{D_{\ell,t}^{m,j}})R_{t+\ell}$ is nonnegative and increases as
$\gamma_w$ decreases.
\endgroup

{\color{black}
\section{Experiment Hyperparameters}
\label{app:experiment-hyperparameters}

\subsection{Models, LaMAS Configurations, and Token Budgets}

The mathematics runs use Qwen2.5-1.5B-Instruct with the DeepScaleR
training split; the coding runs use Qwen2.5-3B-Instruct with the DeepCoder
Preview training split. The actor, reference model, critic, and generation
server are initialized from the same domain-specific base model.
For both DeepScaleR and DeepCoder, the evaluation examples
are drawn from separately held-out evaluation subsets and do not overlap with
the corresponding training examples.

All evaluated LaMAS configurations use a sequential graph. The specialized
\Duo{} configuration uses Planner~$\rightarrow$~Solver. \Trio{} adds a
Verifier, and \Quad{} inserts a Reflector before the Verifier. Their anonymous
counterparts retain the graph and use position-indexed generic task-solving
prompts. For \Duo{}, \Trio{}, and \Quad{}, the configured
per-agent generation limits are respectively 12,288, 8,192, and 6,144 new
tokens for mathematics, and 13,824, 9,216, and 6,912 for coding. These are
request-level caps for each agent. The 24,576-token server context covers the
problem, available conversation history, and the current completion, so the
realized completion can be shorter than its configured cap.

The mathematics input cap is 1,024 tokens, with training and online-validation
batch sizes of 256 and 64. The coding input cap is 4,096 tokens, and both
batch sizes are 256. One rollout is sampled per training prompt at temperature
1.0; sampling is non-greedy.
All named and anonymous \Duo{}, \Trio{}, and \Quad{}
configurations use one PPO minibatch for the actor and one for the critic.

\subsection{Optimization and Multi-Agent Settings}

The actor and centralized critic use AdamW with learning rate
$5\times10^{-6}$, weight decay 0.01, gradient clipping at 1.0,
$(\beta_1,\beta_2)=(0.9,0.999)$, and $\epsilon=10^{-8}$. The actor uses cosine
learning-rate decay, a 0.001 warmup proportion, and a final-to-initial
learning-rate ratio of 0.05. The critic uses a constant schedule, the same 0.001
warmup proportion, and no terminal learning-rate decay. PPO uses a clipping parameter of 0.2. The KL coefficient is 0.1 with
the $k_3$ estimator. Log probabilities are recomputed, the decoupled loss is
enabled, and behavior-importance weights are capped at 5. Rewards and
advantages are normalized at batch level. The actor and critic use bfloat16,
dropout is disabled, gradient checkpointing is enabled, and each actor/critic
microbatch is capped at 32,768 tokens. The generation-server context length is
24,576 tokens.

Every policy and critic adapter uses LoRA rank 32, scaling factor 16, and all
linear target modules. With policy parameter sharing, all roles update
one policy adapter. Without policy parameter sharing, each role updates its own
policy adapter. Both conditions retain one centralized critic. All agents
receive the shared team reward, which is assigned equally to the role-level
actions in the sequential graph.

\subsection{Evaluation Protocol}

During training, online evaluation uses non-greedy decoding at temperature
0.6. The mathematics monitoring procedure draws 16 completions per AIME2024
problem and one completion per problem for MATH500, OlympiadBench, and Minerva
Math; the coding monitor draws one completion per instance.

Each final-table evaluation record identifies the evaluated trained
checkpoint by its global step; scores from different checkpoint steps are not
pooled. Each benchmark item is decoded once under each of five stochastic
evaluation seeds at temperature 0.6. The
procedure first averages rewards over all items for each decoding seed and then
reports the mean across the five seed-level averages. These seeds quantify
decoding variation for one trained checkpoint rather than variation across five
independent training runs. Mathematics uses binary correctness; coding uses the
benchmark-specific execution reward, reported as pass rate or score. Base-model
results use the same LaMAS prompts and decoding protocol without loading a
trained LoRA adapter.

\subsection{System Prompts}
\label{app:system-prompts}
\newcommand{\promptcode}[1]{{\normalfont\texttt{#1}}}

The benchmark problem is supplied as the initial user message. Each agent
receives its own system prompt together with the available conversation history
generated by preceding agents. Before a noninitial agent acts, the sequential
configuration inserts the user turn
\texttt{Now it is \{role\}'s turn. \{role description\}}, where \texttt{role}
is the configured role name and \texttt{role description} is the corresponding
fixed description listed below. The following are the system prompts used in
the reported runs; line breaks are editorial only. In the coding prompts, the
literal \texttt{...} between the Python fences is part of the prompt and marks
where generated code should appear.

The transition descriptions are also fixed. For mathematics they are:
Solver, ``Executes the plan and produces the final answer.''; \Trio{} Verifier,
``Verifies or corrects the solver's answer.''; \Quad{} Reflector, ``Reviews the
solution and identifies errors or gaps.''; and \Quad{} Verifier, ``Verifies or
corrects the solver's answer after reflection.'' For coding they are: Solver,
``Implements the plan as Python code.''; \Trio{} Verifier, ``Verifies and
finalizes the code.''; \Quad{} Reflector, ``Reflects on the code and identifies
issues.''; and \Quad{} Verifier, ``Produces the final verified code.'' Each
anonymous agent uses ``The \{ordinal\} agent in the sequential pipeline.''

\subsubsection{Mathematics}

\paragraph{Planner ({\normalfont\Duo{}}/{\normalfont\Trio{}}/{\normalfont\Quad{}}).}
\emph{You are a math problem planner. Given a math problem, break it down into
clear, numbered steps. Do NOT solve the problem---only produce the solution
plan.}

\paragraph{Solver ({\normalfont\Duo{}}/{\normalfont\Trio{}}/{\normalfont\Quad{}}).}
\emph{You are a math solver. Follow the plan provided in the conversation and
solve the problem step by step. Put your final answer inside
\promptcode{\textbackslash boxed\{\}}.}

\paragraph{Verifier ({\normalfont\Trio{}}).}
\emph{You are a math verifier. Check the solution provided in the conversation
for correctness. If the answer is correct, restate it inside
\promptcode{\textbackslash boxed\{\}}. If it is wrong, provide the corrected answer
inside \promptcode{\textbackslash boxed\{\}}.}

\paragraph{Reflector ({\normalfont\Quad{}}).}
\emph{You are a math solution reviewer. Read the plan and solution provided in
the conversation. Identify any errors, gaps, or unjustified steps. If the
solution is correct, say so briefly. If there are mistakes, explain them
clearly.}

\paragraph{Verifier ({\normalfont\Quad{}}).}
\emph{You are a math verifier. Taking into account the plan, solution, and
review provided in the conversation, produce the final correct answer inside
\promptcode{\textbackslash boxed\{\}}. If the solution was correct, restate it; if
it was wrong, provide the corrected answer.}

\paragraph{Anonymous agents.}
For anonymous agent $A_i$, the ordinal corresponding to its position is
inserted into the following common template: \emph{You are the
\{ordinal\} agent in a multi-agent math problem solving team. Work on the math
problem based on the conversation so far. Put your final answer inside
\promptcode{\textbackslash boxed\{\}}.}

\subsubsection{Coding}

\paragraph{Planner ({\normalfont\Duo{}}/{\normalfont\Trio{}}/{\normalfont\Quad{}}).}
\emph{You are a programming problem planner. Given a coding problem, analyze
the requirements, identify edge cases, and produce a clear step-by-step
algorithm plan. Do NOT write code---only produce the solution plan with data
structures and approach.}

\paragraph{Solver ({\normalfont\Duo{}}/{\normalfont\Trio{}}/{\normalfont\Quad{}}).}
\emph{You are an expert Python programmer. Follow the plan provided in the
conversation and implement a correct Python solution. Enclose your code within
\promptcode{\textasciigrave\textasciigrave\textasciigrave python ...
\textasciigrave\textasciigrave\textasciigrave} delimiters.}

\paragraph{Verifier ({\normalfont\Trio{}}).}
\emph{You are a code verifier. Check the code provided in the conversation for
correctness, edge cases, and potential bugs. If the code is correct, restate it
within \promptcode{\textasciigrave\textasciigrave\textasciigrave python ...
\textasciigrave\textasciigrave\textasciigrave} delimiters. If it has bugs,
provide the corrected version within
\promptcode{\textasciigrave\textasciigrave\textasciigrave python ...
\textasciigrave\textasciigrave\textasciigrave} delimiters.}

\paragraph{Reflector ({\normalfont\Quad{}}).}
\emph{You are a code reflector. Review the code provided in the conversation
for correctness, edge cases, and potential bugs. Provide your analysis and
suggest specific fixes if needed.}

\paragraph{Verifier ({\normalfont\Quad{}}).}
\emph{You are a code verifier. Based on the code and reflection, produce the
final Python solution. If the code is correct, restate it within
\promptcode{\textasciigrave\textasciigrave\textasciigrave python ...
\textasciigrave\textasciigrave\textasciigrave} delimiters. If it has bugs,
provide the corrected version within
\promptcode{\textasciigrave\textasciigrave\textasciigrave python ...
\textasciigrave\textasciigrave\textasciigrave} delimiters.}

\paragraph{Anonymous agents.}
For anonymous agent $A_i$, the ordinal corresponding to its position is
inserted into the following common template: \emph{You are the
\{ordinal\} agent in a multi-agent coding problem solving team. Work on the
coding problem based on the conversation so far. Enclose your final solution
within \promptcode{\textasciigrave\textasciigrave\textasciigrave python ...
\textasciigrave\textasciigrave\textasciigrave} delimiters.}
}

\end{document}